\documentclass[preprint,11pt]{elsarticle}
\usepackage{color}
\usepackage{algorithm}
\usepackage{algorithmic}
\usepackage{booktabs}
\usepackage{ulem}
\usepackage{ifthen}

\definecolor{gray}{rgb}{0.5,0.5,0.5} 
\definecolor{green}{rgb}{0, 0.4, 0} 
\definecolor{orange}{rgb}{1, 0.5, 0} 	
\definecolor{mahogany}{rgb}{0.75, 0.25, 0.0}
\definecolor{purple}{rgb}{0.6, 0, 0.6}

\newcommand{\ignore}[1]{}

\newboolean{revising}
\setboolean{revising}{false}
\ifthenelse{\boolean{revising}}
{
	\newcommand{\mate}[1]{\textcolor{purple}{#1}}
	\newcommand{\mateComment}[1]{\textcolor{red}{#1}}
	\newcommand{\michael}[1]{\textcolor{green}{#1}}
	
	\newcommand{\mateignore}[1]{\textcolor{purple}{\sout{#1}}}
	\newcommand{\michaelignore}[1]{\textcolor{green}{\sout{#1}}}
	\newcommand{\ameanignore}[1]{\textcolor{orange}{\sout{#1}}}
} {
	\newcommand{\mate}[1]{#1}	
	\newcommand{\mateComment}[1]{}	
	\newcommand{\michael}[1]{#1}
	
	\newcommand{\michaelignore}[1]{}
	\newcommand{\mateignore}[1]{}
	\newcommand{\ameanignore}[1]{}
}

\usepackage{picins}
\usepackage[hyphens]{url}
\usepackage[hyperindex,breaklinks]{hyperref}
\usepackage{breakurl}

\journal{Ad Hoc Networks}

\usepackage{amssymb}

\biboptions{square}

\usepackage[figuresright]{rotating}

\newcommand{\textblue}[1]{{#1}}
\newcommand{\textred}[1]{{#1}}

\usepackage{epstopdf}
\graphicspath{{pic/}}
\usepackage{tabularx}
\usepackage[cmex10]{amsmath}
\usepackage{subfigure}

\begin{document}

\begin{frontmatter}




\title{Scooter-to-X Communications: Antenna Placement, Human Body Shadowing, and Channel Modeling}


\author[rvt]{Hao-Min Lin}
\ead{d00922003@csie.ntu.edu.tw}
\author[rvt]{Hsin-Mu Tsai\corref{mycorrespondingauthor}}
\cortext[mycorrespondingauthor]{Corresponding author}
\ead{hsinmu@csie.ntu.edu.tw}
\author[els]{Mate Boban}
\ead{mate.boban@huawei.com}

\address[rvt]{Department of Computer Science and Information Engineering,\\ National Taiwan University, Taipei, Taiwan.}
\address[els]{Huawei European Research Center, Munich, Germany.}

\begin{abstract}
In countries such as Taiwan, with a high percentage of scooters, scooter-related accidents are responsible for most injuries and deaths of all traffic accidents. 
One viable approach to reduce the number of accidents is to utilize short-range wireless communications between the scooter and other vehicles. This would help neighboring vehicles to detect the scooter and vice-versa, thus reducing the probability of a collision. 
In this paper, we perform extensive measurements to characterize communication links between a scooter and other vehicles. Our results suggest that, when the line-of-sight propagation path is blocked by the body of the scooter driver (and possibly also a passenger), shadowing of the human body results in significant signal attenuation, ranging from 9 to 18 dB on average, presenting challenging channel characteristics unique to scooters. In addition, we perform simulations, which show that it is imperative to incorporate the body shadowing effect to obtain realistic simulation results. We also develop a model to determine whether human body shadowing is in effect, given the relative positions of the transmitter and the receiver.
\end{abstract}


\end{frontmatter}



\section{Introduction}
\label{sec:intro}
In recent years, various safety systems have gradually become standard features in new car models. Examples include Anti-lock Breaking System (ABS), Electronic Stability Control (ESC) system, lane departure warning (LDW) system, etc. Despite this, in countries with high proportion of scooters\footnote{Scooters refer to a type of motorcycle that has an automatic transmission and a step-through frame. In this paper, we use the term ``scooter'' to refer to both scooters and other types of motorcycles, since scooters are the most popular type of motorcycle in most south eastern Asian countries.} (e.g., Taiwan) the number of accidents, injuries and deaths caused by the accidents have been steadily increasing over the past decade~\cite{IOT-2011}. Closer investigation reveals that the increase is mainly caused by accidents involving scooters; on average, the numbers have been increasing by 10\% every year and they are responsible for more than 80\% - 90\% of the deaths and the injuries~\cite{IOT-2011}.

In countries such as Vietnam, Malaysia, and Taiwan, close to 70\% of the registered vehicles are scooters, 
\textblue{sold for approximately \$2,000 USD (i.e., an order of magnitude less than personal vehicles), thus requiring the development of new and cost effective safety systems designed for scooters.}
Wireless communications between vehicles -- both cars and scooters -- are the key to enabling safety systems, as they provide the necessary means for cooperation and information sharing and could eliminate the need for incorporating costly sensors in every scooter. If the location and the status of the scooters and the cars are periodically broadcast, neighboring vehicles would have better awareness of the surrounding environment and the probability of a collision would be greatly reduced~\cite{papadimitratos2009vehicular}.

In the past decade, many protocols and systems have been proposed to enable Vehicular Ad-Hoc Networks (VANETs). Radios complying to IEEE 802.11p~\cite{IEEE80211pstd} standard, commonly referred to as Dedicated Short Range Communications (DSRC), are the most prominent technology for enabling VANETs. 
\textblue{On the other hand, 2.4 GHz communication technologies such as ZigBee (IEEE 802.15.4), WiFi (IEEE 802.11b/g/n), and Bluetooth have been widely used for a variety of applications, and present themselves as alternative and cost-effective solutions to implement scooter-to-X communications.} 
Furthermore, past studies have shown that attenuation caused by both shadowing in VANETs and shadowing due to pedestrians in indoor environments exhibit similar behaviors at 2.4 GHz and at 5.9 GHz~\cite{boban11, Villanese2000}.
\textblue{Therefore, in this paper we evaluate the ability of radio technologies using the 2.4 GHz ISM spectrum to form vehicular networks with scooters and cars.}

To incorporate a wireless radio on a scooter, many design questions remain to be answered. The antenna on a car is usually mounted at the highest location, which is usually the roof. Kaul et al. have shown the center of the roof to be the overall best position for the antenna~\cite{SECON07}. However, on a scooter, it is not obvious where the optimal antenna location is. The propagation path between the transmitting and receiving antennas on scooters could be blocked by the driver and the additional passenger on the scooters, resulting in additional path loss and received power variations. 
In this paper, we present results of measurements between scooters on the road utilizing IEEE 802.15.4 radios (Zigbee)~\cite{Zigbee} \textred{operating} at 2.4 GHz, which provide insights and guidelines for these design questions. Specifically, our main contributions are the following:

\begin{itemize}
\item To the best of our knowledge, we perform the first extensive measurement study aimed at characterizing the scooter-to-X links.  
\item We empirically identify the optimal antenna location with the minimum signal attenuation for a 2.4 GHz radio on the scooter.
\item By performing extensive measurements, we analyze in detail the additional signal attenuation caused by the driver and the passenger of the scooter.
\item We propose a model that incorporates the additional signal attenuation due to a scooter driver and passengers, based on the relative locations of the transmitting and receiving vehicles. The proposed model can be used for simulations or theoretical studies.
\item We carry out simulations to show that it is crucial to consider the additional signal attenuation caused by the driver and the passenger of the scooter.
\end{itemize}

\section{Related Work}
\label{sec:related_work}

Our study has three goals: 1) find the best location for mounting antenna on a scooter; 2) understand the impact of human body shadowing on scooter-to-X links; and 3) understand the propagation characteristics of scooter-to-X links when scooters are located in different locations on the road (i.e., when they are in different lanes).

With regards to 1), the work on antenna placement on scooters has been virtually nonexistent. With regards to personal cars, Kaul et al.~\cite{SECON07} analyzed the performance of six different antenna placements on the rooftop of a personal vehicle. The results showed that antenna locations lead to 25-30\% difference in cumulative link packet error rate. The antenna location with the best overall performance was found to be in the center of the roof. 
The same study also showed 10-15\% of packet error rate reduction when multiple receive antennas and packet level diversity techniques are used.
Oh et al.~\cite{oh09} explored vertical diversity by placing antennas on a rooftop of a personal vehicle and inside the passenger cabin. The vertical diversity was shown to increase the effective communication range by 100 meters in certain scenarios. Similarly, study by Boban et al.~\cite{bobanTVRTMC} showed that antennas positioned on rooftops of vans and trucks, due to their taller position, result in less shadowing and provide increased packet delivery and communication range when compared to antennas on shorter, personal vehicles. 

The studies above have experimentally shown that mounting the antenna on vehicle roof results in best overall communication performance. 
Following the logic of these findings, to improve the scooter-to-X link quality, the best position for the antenna on a scooter would be on the tallest possible position. 
However, in case of scooters, mounting antenna on the tallest position, driver's or passenger's helmet, is not practical.
For these reasons, we investigate fixed antenna positions on different locations on the scooter.

Similar to antenna placement, the impact of human body shadowing in scooter-to-X links is virtually unexplored. Therefore, we refer to human body shadowing in other environments to provide insight for our study of scooter-to-X links.  
Yamaura~\cite{HumanTorso} measured 10~dB attenuation at 2~GHz frequency due to human body shadowing. Similarly, Huo et al. in~\cite{HumanAct} studied the effect of human shadowing on 2.4~GHz radio in home environment, and reported maximum 20 dB attenuation when the height of transmitter and receiver is 70 cm. 

In terms of modeling shadowing (i.e., signal variation due to objects larger than the carrier wavelength) for \textred{V2V} communication, Gallagher et al.~\cite{gallagher06} analyzed the impact of vehicle shadowing on packet reception, throughput, and communication range. A statistical human body shadowing model for indoor environments has been developed by Obayashi and Zander~\cite{obayashi1998body}. The model can be used to estimate the shadowing using specific indoor layouts. Boban et al.~\cite{boban11} analyzed the impact of vehicles as obstacles and proposed a propagation model that incorporates vehicular obstructions. Abbas et al.~\cite{abbas12} performed V2V measurements to quantify the duration of shadowing due to vehicles; based on measurements, the authors designed a stochastic model for highways that incorporates vehicular obstructions. 
GEMV$^2$ is an efficient geometry-based propagation model for V2V communications proposed by Boban et al.~\cite{boban14TVT}, which explicitly accounts for surrounding objects (buildings, foliage and other vehicles). The model considers three V2V links categories, depending on the LOS conditions between transmitter and receiver, to deterministically calculate large-scale signal variations (i.e., path-loss and shadowing): i) LOS links; ii) non-LOS links obstructed by other vehicles~\cite{boban11}; and iii) non-LOS links whose LOS is obstructed by buildings or foliage. GEMV$^2$ is implemented in a freely available V2V propagation model and simulation framework available at \url{http://vehicle2x.net}. 
For more details on propagation modeling as applied to personal and commercial vehicles, Matolak~\cite{matolak13} and Viriyasitavat et al.~\cite{viriyasitavat2015vehicular} provide a comprehensive overview of recent developments.

The studies described above, along with numerous others focusing on V2V communications have been performed, wherein vehicles are assumed to be either personal cars or large vehicles such as buses and trucks. To the best of our knowledge, there have been no comprehensive studies investigating in detail the properties of scooter-to-X links. The frame structure, driving/riding position, dimension and the position of mounted antenna are considerably different on  scooters when compared to other types of vehicles. 
Therefore, to enable development and realistic evaluation of protocols and applications related to scooter-to-X communication, we first characterize in detail the scooter-to-X channel by performing extensive measurements that evaluate the effect of both antenna placement and body shadowing.
Then, we develop a model that incorporates these phenomena into the scooter-to-X channel model.

\section{2.4 GHz Scooter-to-Scooter Link Measurements}
\label{sec:link_measurement}


%

\subsection{Experimental Setup}
\label{subsec:exp_setup}
\qquad
In this subsection, we present the general parameters and the settings of the carried out experiments.
We use Zigduino~\cite{Zigduino} as our main hardware platform. The Zigduino's on-board microprocessor, TI ATMega128RFA1~\cite{atmel}, combines a 8-bit 16 MHz processor and an IEEE 802.15.4~\cite{Zigbee} radio in a single ship. nano-RK~\cite{nanoRK} is used to implement all firmware for the experiments. An 8 dBi omni-directional rubber duck antenna (approximately 17.5-cm long) is attached to each Zigduino board, \michael{and the antennas are placed vertical to the ground}.
The transmitter is configured to transmit with a transmission power of 3.3 dBm (2.14 mW), which is the maximum transmission power setting on Zigduino. Channel 26 (2.480 GHz) is used as the operating channel and frequency for all experiments. The ATmega128RFA1 provides automatic energy detection (ED) with a 128 $\mu$s measurement duration, after detection of a valid synchronization header (SHR) of an incoming frame. \michael{This measurement estimates the received signal power within the bandwidth of an IEEE 802.15.4 channel~\cite{atmel}, which is 2 MHz.} We use ED level to represent the RSSI value and its minimum value is -90 dBm. Figure 9-18 in~\cite{atmel} shows that the measurement is fairly accurate with only 1-2 dB of error when the received power is in the range of -50 to -90 dBm.
\michaelignore{, shown as the dotted line in Figure~\ref{fig:RSSI_PRR}. We collected the experimental data and each data point represents the reception rate of 10 packets in Figure~\ref{fig:RSSI_PRR}, which shows a drop of packet reception rate to 0 at -90 dBm and verifies the stated minimum RSSI value, i.e., the receive sensitivity threshold.} 

For each transmitter-receiver pair, the transmitter is configured to transmit one packet per 100 ms for 5000 consecutive packets. Each packet is 12 bytes in length with a sequence number. ATmega128RFA1's 802.15.4 radio adds a hardware calculated CRC checksum to each transmitted packet and all received packets are checked against the attached checksum. If the calculated checksum based on the received packet content is different from the attached checksum, the received packet is regarded as erroneous.

In order to minimize the impact of WiFi interference, multi-path propagation, and other factors, we performed all experiment in an empty and closed road segment in Da-Jia Riverside Park, Taipei, Taiwan (see Figure~\ref{fig:ant_loc}(b)). We used YAMAHA Cygnus-X 125 (2011 year model) for both the transmitter and the receiver. The dimensions of scooter are 185.5 x 68.5 x 115 cm (Length x Width x Height).

\subsection{Antenna location comparison: experiments without human body shadowing}
\label{subsec:ant_loc}
%
%

\begin{figure}[t]
\centering
\subfigure[Tested Antenna Locations]{\includegraphics[width=0.45\textwidth]{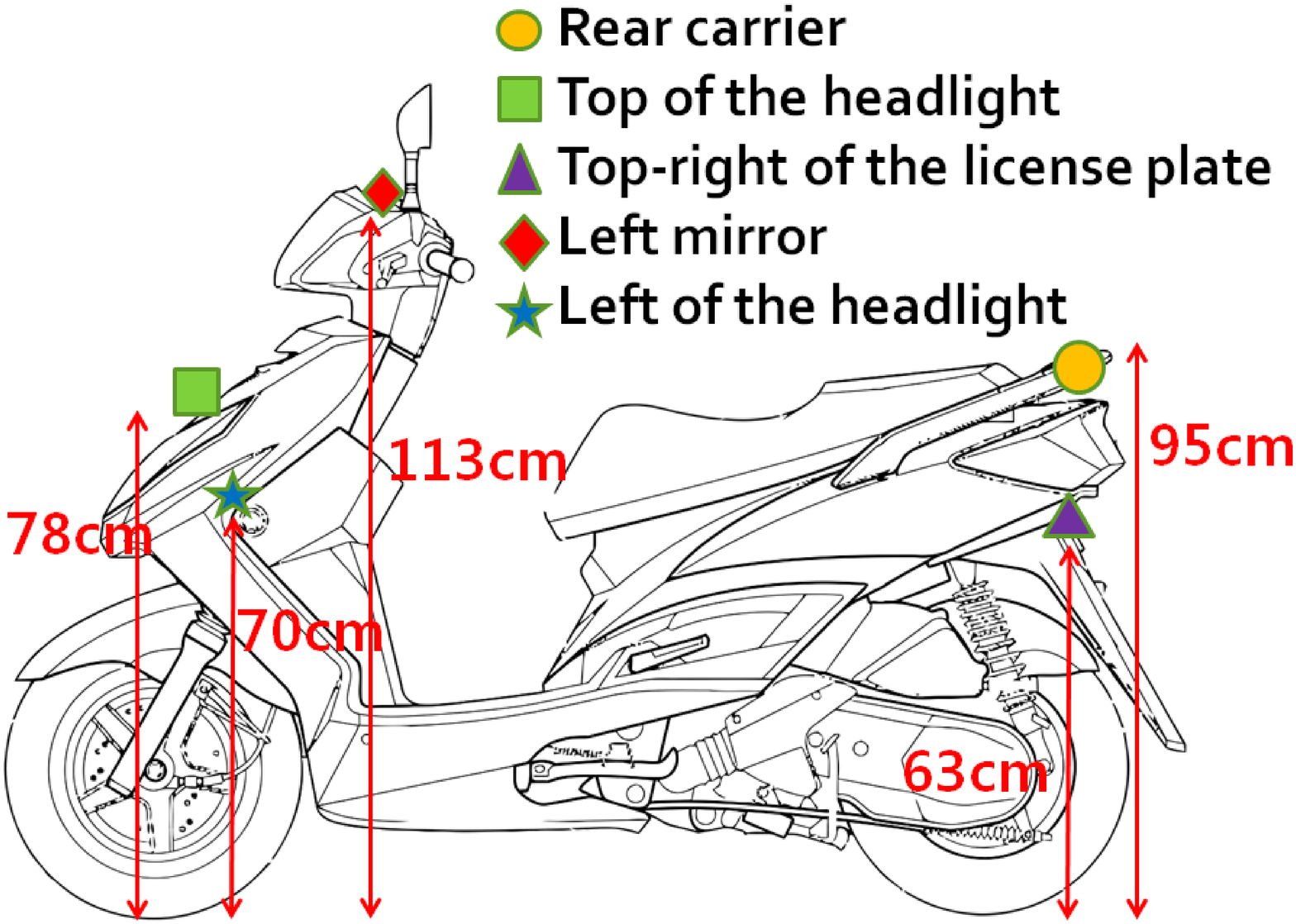}}
\subfigure[Transmitting and Receiving Locations]{
\includegraphics[width=0.45\textwidth]{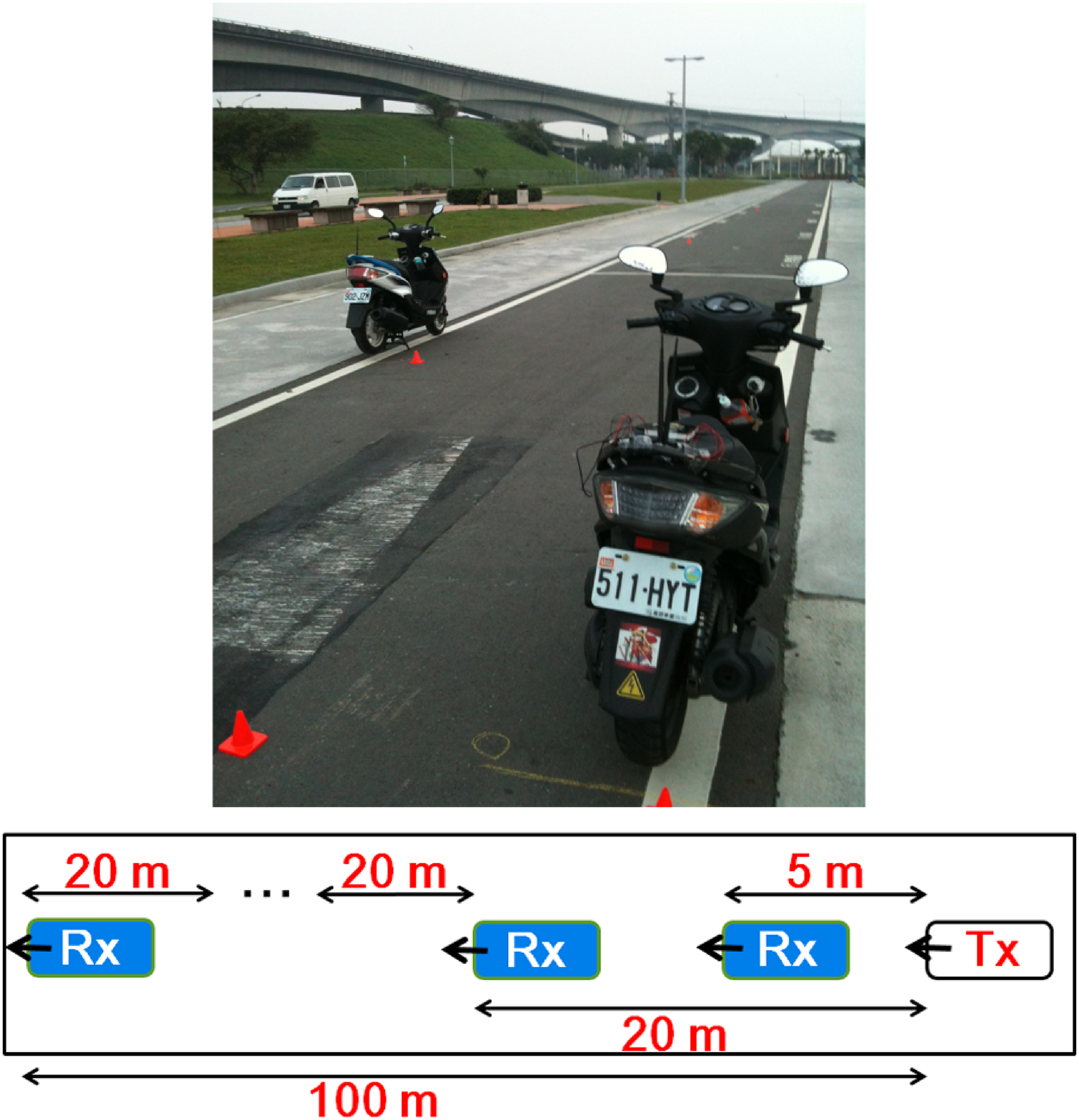}
}
\caption{The Experimental Setup}
\label{fig:ant_loc}
\end{figure}

\begin{figure}[t]
\centering
\includegraphics[width=0.6\textwidth]{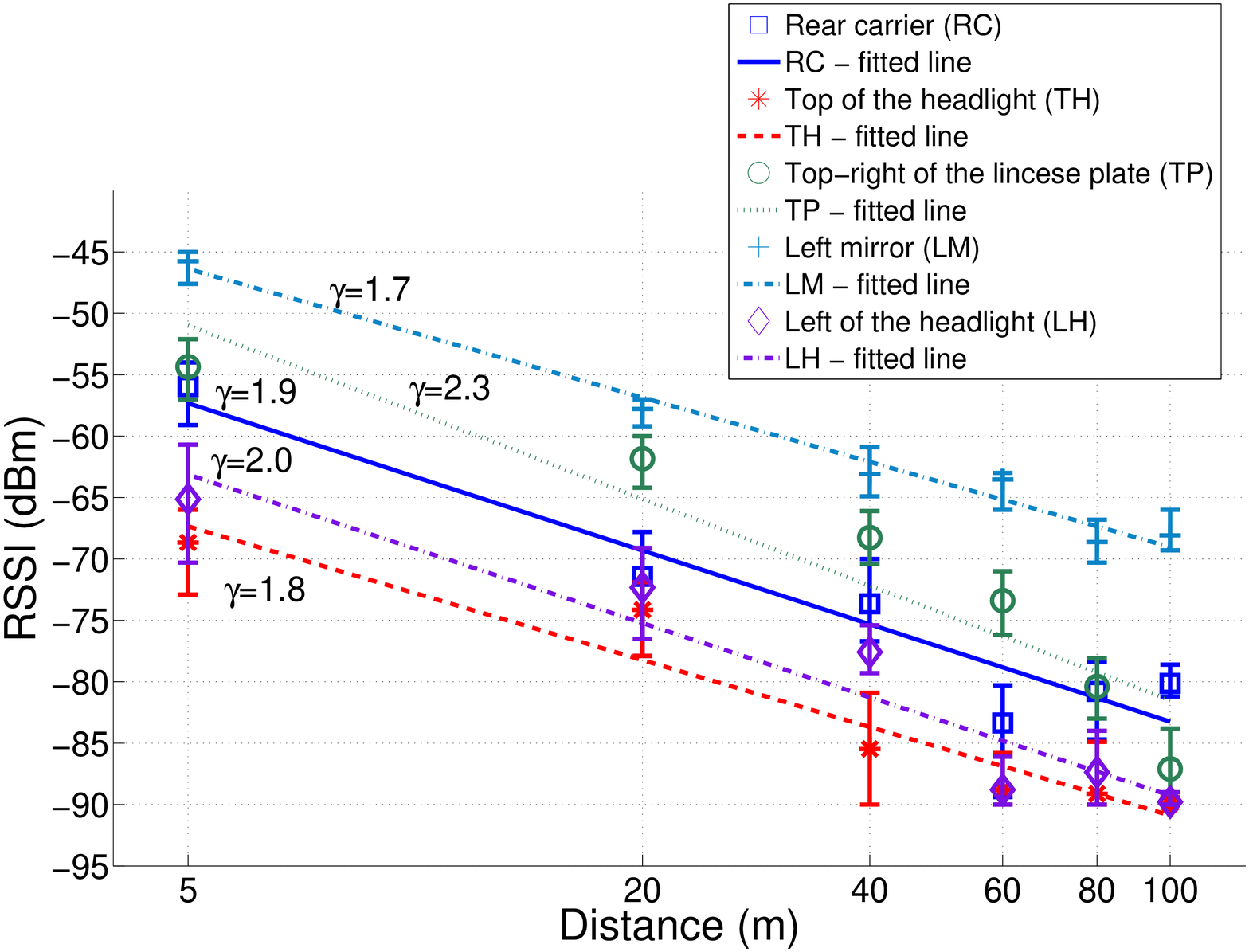}
\caption{Comparison of RSSI when different antenna locations are used. The error bars mark the 5th and the 95th percentiles as well as the mean values of all obtained RSSI.}
\label{fig:AntennaLocationCompare1}
\end{figure}

The goal of this experiment is to determine possible optimal antenna location on the scooter, which introduces the least path loss on average, in cases when there is no driver or passenger on the scooter. This serves as a preliminary study to understand how sensitive the path loss is to the antenna location.
In this experiment, we compared five different antenna locations on a scooter (see Figure~\ref{fig:ant_loc}(a)). 

\begin{enumerate}
\item Rear carrier: The rear carrier is the highest location of the tail of a scooter (95 cm from the ground).
\item Top of the headlight: The top of the headlight is the frontmost location of the scooter (78 cm from the ground) which is neither blocking the headlight nor affected by the front wheel.
\item Top-right of the license plate: The top-right of the license plate is the most common location for a short wave radio antenna (for voice communications). The location is 63 cm from the ground.
\item Left mirror: The left mirror (or the right mirror) is the highest location of a scooter (113 cm from the ground) and does not affect the vision of a driver.
\item Left of the headlight: Located on the left of the headlight (70 cm from the ground).
\end{enumerate}

For this experiment, we place two scooters in the same lane; the transmitter is on the front scooter and the receiver is on the rear scooter, \michael{as shown in Figure~\ref{fig:ant_loc}(b).} 
For each run, the transmitter sends 5000 packets.
Figure~\ref{fig:AntennaLocationCompare1} shows average RSSI versus distance in log scale between the transmitter and the receiver for each antenna location. The lines in Figure~\ref{fig:AntennaLocationCompare1} represent the fitting results of the log-distance path loss model given the RSSI values collected from the experiments.

The log-distance path loss model is given by
\begin{equation}
\begin{aligned}
P_L(d)[dB] &= \bar{P_L}(d_0) + 10\gamma log_{10}\frac{d}{d_0} + X_\sigma \\
&= 10\gamma log_{10}d + (\bar{P_L}(d_0) - 10\gamma log_{10}d_0 + X_\sigma) \\
&= 10\gamma log_{10}d + X'_\sigma
\end{aligned}
\label{eq:pathloss}
\end{equation}
where $\bar{P_L}(d_0)$ represents the path loss when the distance between the transmitter and the receiver (T-R distance) is $d_0$ (a reference distance), $\gamma$ is the path loss exponent, and $X_\sigma$ describes the random shadowing effects. $X'_\sigma$ is used to represent all the other effects besides T-R distance. 

As can be seen in Figure~\ref{fig:AntennaLocationCompare1}, all fitted lines are approximately parallel, except for the scenario where the antenna was mounted on the top-right of the license plate. In this location, the antenna is at the corner of the scooter; when the two scooters are placed at locations with different distances to each other, a slight inaccuracy in locations would change the channel from a line-of-sight channel (when the front vehicle is slightly to the right) to a non-line-of-sight-channel (when the front vehicle is aligned with the back vehicle or slightly to the left, blocking the propagation path). This results in the irregularities in the path loss measurements for this antenna location shown in the figure.

\begin{table}[t]
\centering
\caption{Comparison of mean attenuation and path loss exponent values of tested antenna locations}
\label{tab:Comparison1}
\resizebox{\textwidth}{!}{
\begin{tabularx}{1.5\textwidth}{@{} | >{\centering}X | >{\centering}X | >{\centering}X | >{\centering}X | >{\centering}X | X<{\centering} | @{}}
\hline
\textbf{Antenna Location}	& \textbf{Rear Carrier}	& \textbf{Top of the Headlight}	& \textbf{Top-Right of the License Plate}	& \textbf{Left Mirror} & \textbf{Left of the Headlight}\\
\hline \hline
 $X'_\sigma$ &-43.4 dB&	-54.7 dB&	-34.6 dB&	-34.2 dB&	-49.1 dB\\
\hline
$\gamma$ &2.0 &1.8 &2.4 &1.7 &2.0\\
\hline
\end{tabularx}
}
\end{table}

\begin{figure}[t]
\centering
\subfigure[Side view, with only the driver]{
\includegraphics[width=0.4\textwidth]{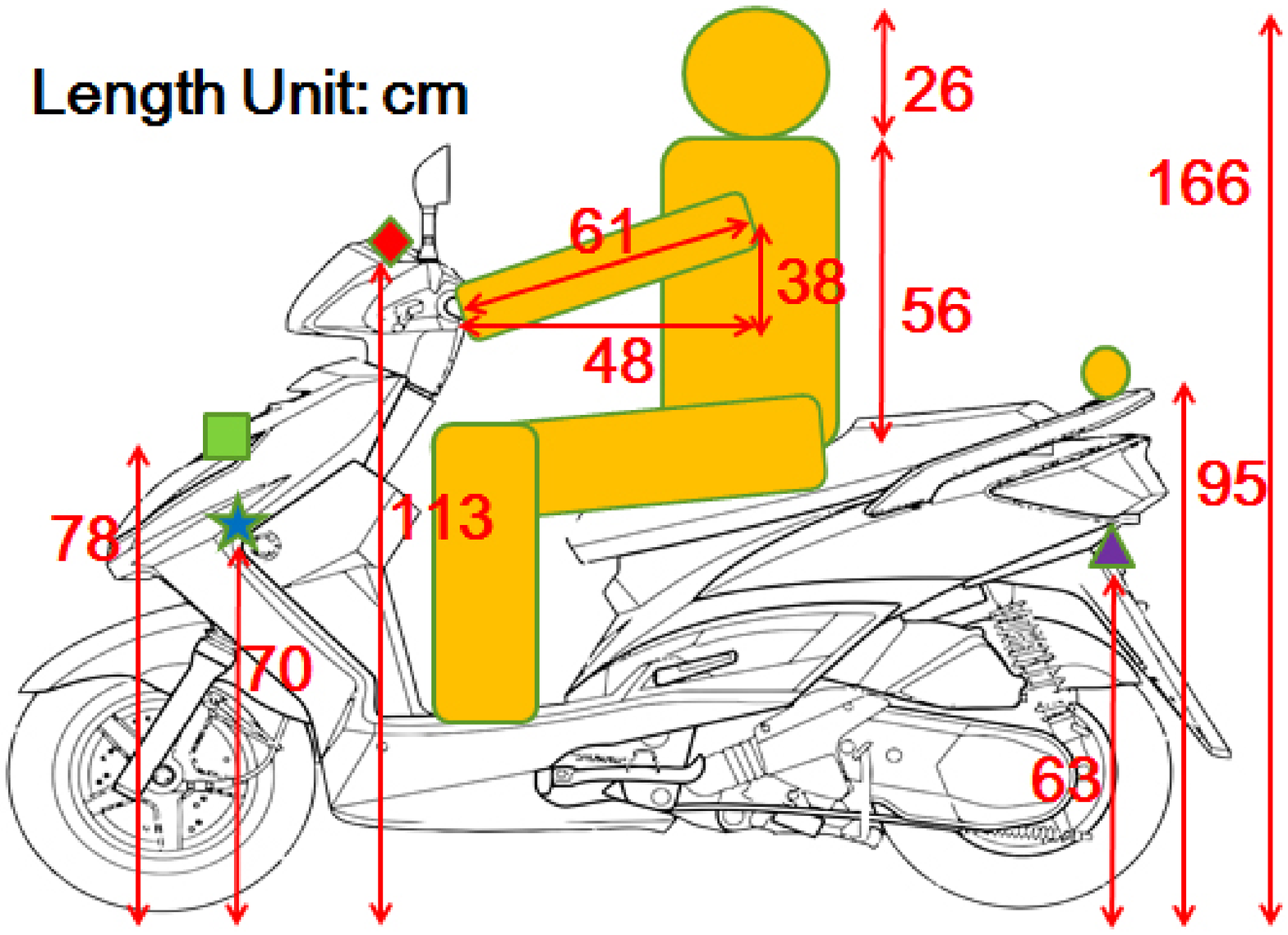}}
\subfigure[Top view, with only the driver]{
\includegraphics[width=0.4\textwidth]{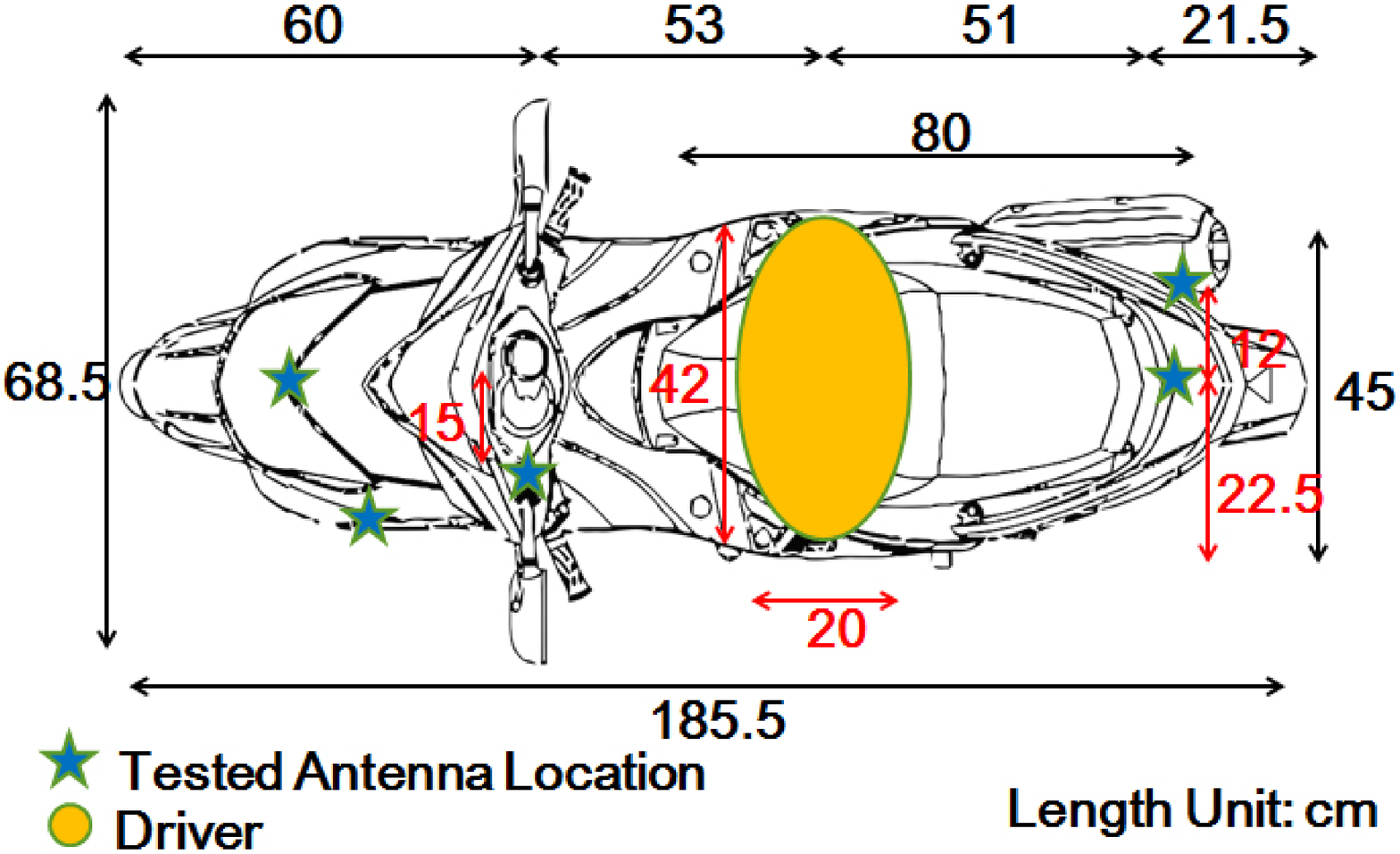}}
\subfigure[Side view, with the driver and the passenger]{
\includegraphics[width=0.4\textwidth]{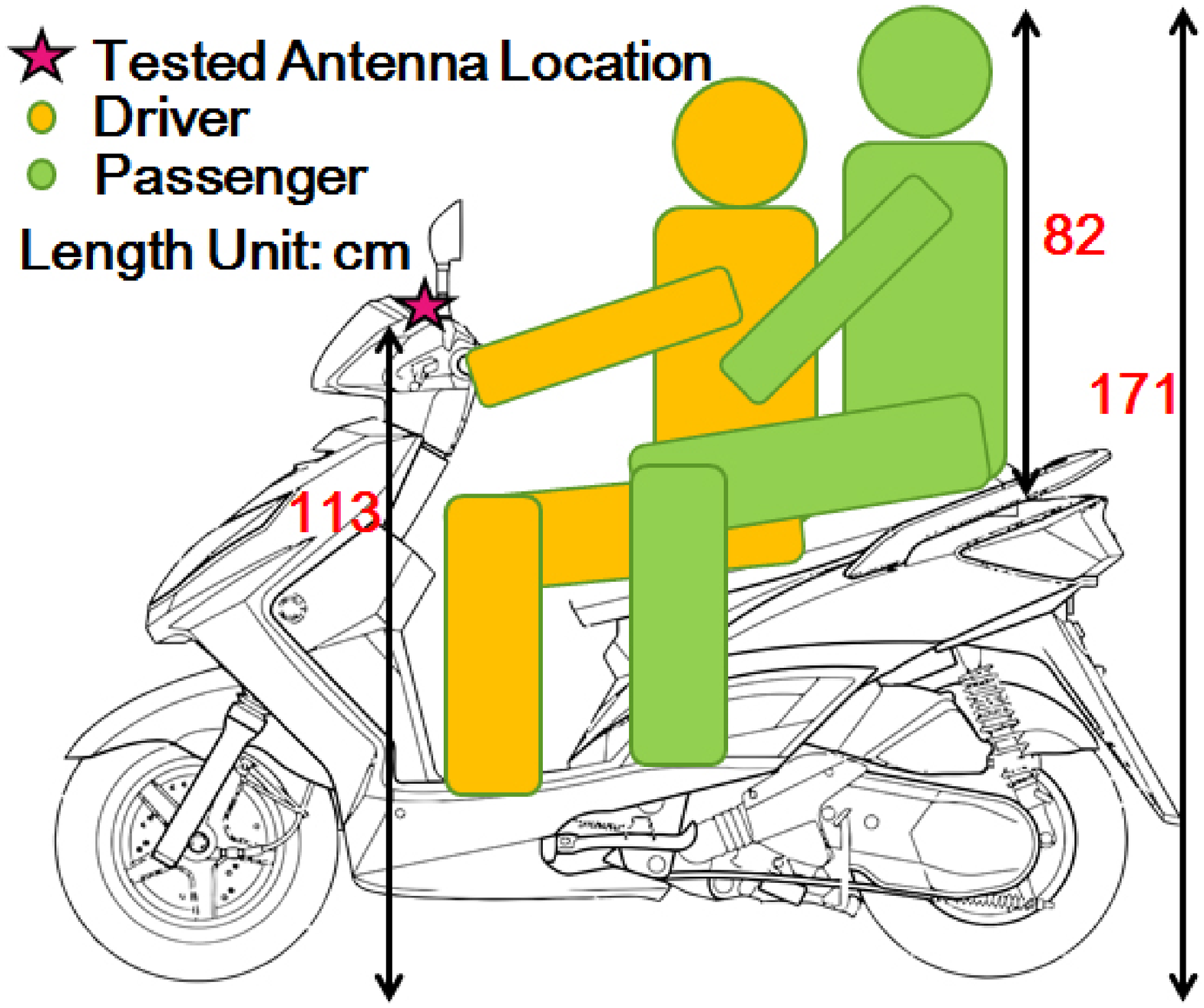}}
\subfigure[Top view, with the driver and the passenger]{
\includegraphics[width=0.4\textwidth]{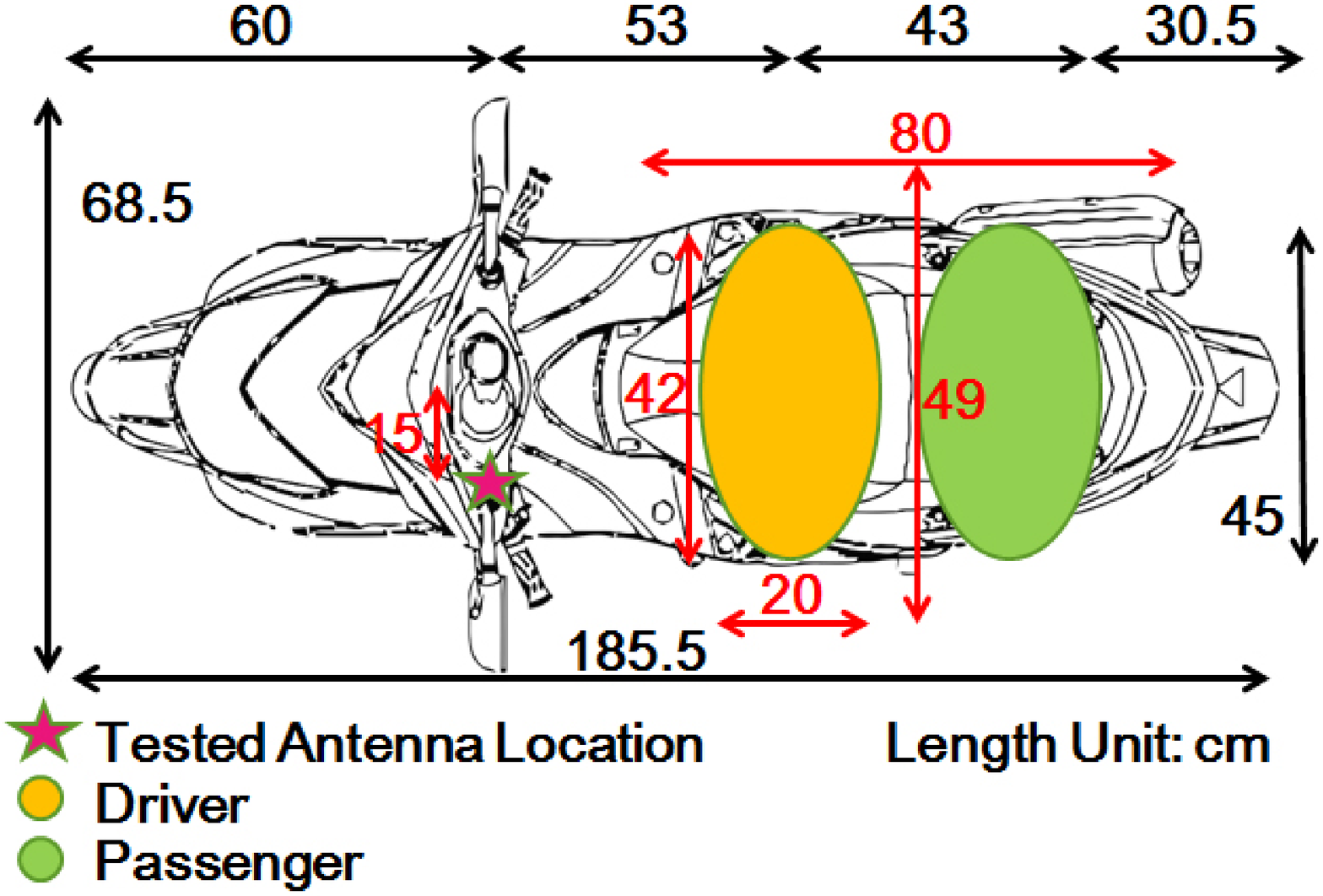}}
\caption{Dimensions of the scooter, the driver, and the passenger}
\label{fig:DriverPassengerDimension}
\end{figure}

\begin{figure}[tp]
\centering
\subfigure[Antenna location: left mirror]{\includegraphics[width=0.48\textwidth]{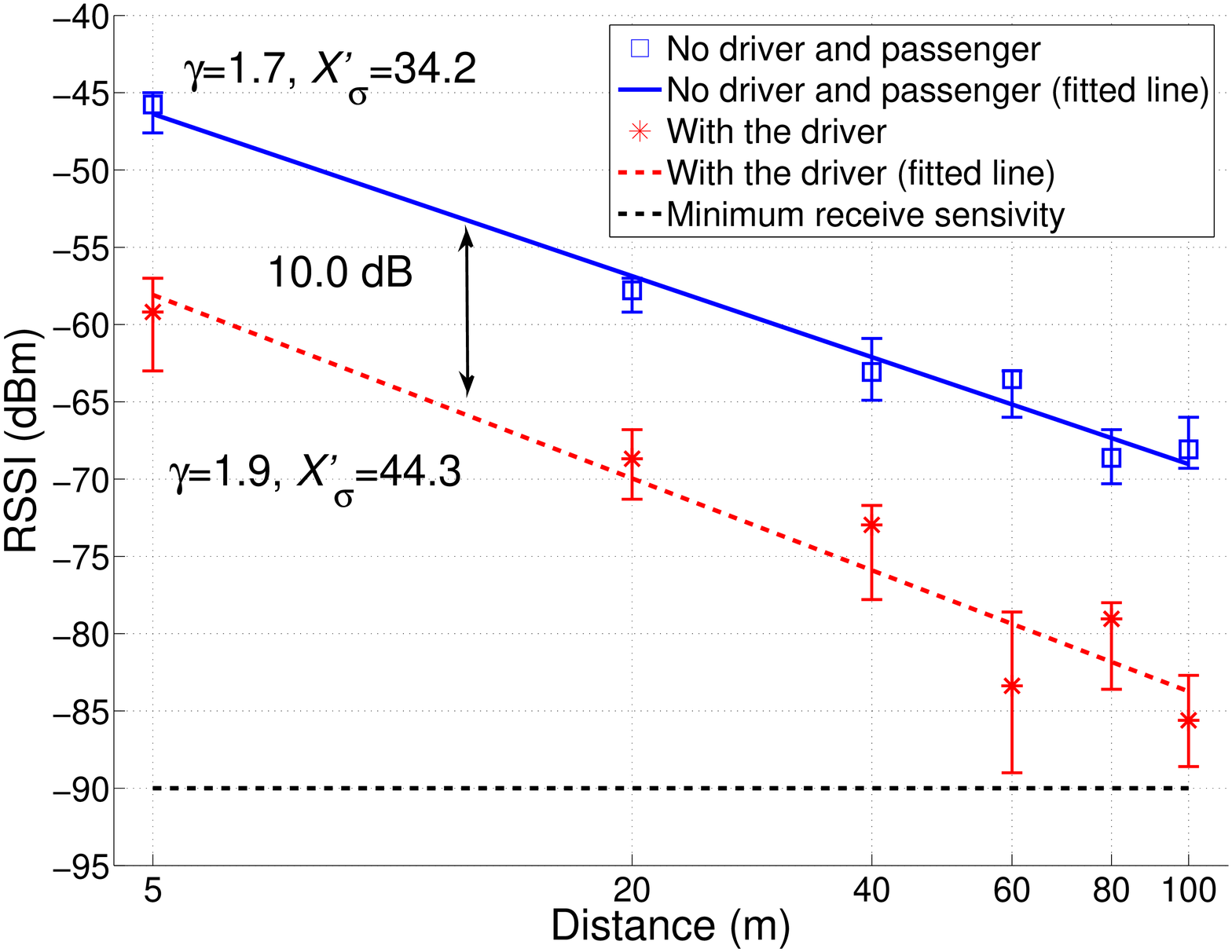}}
\subfigure[Antenna location: top of the head light]{\includegraphics[width=0.48\textwidth]{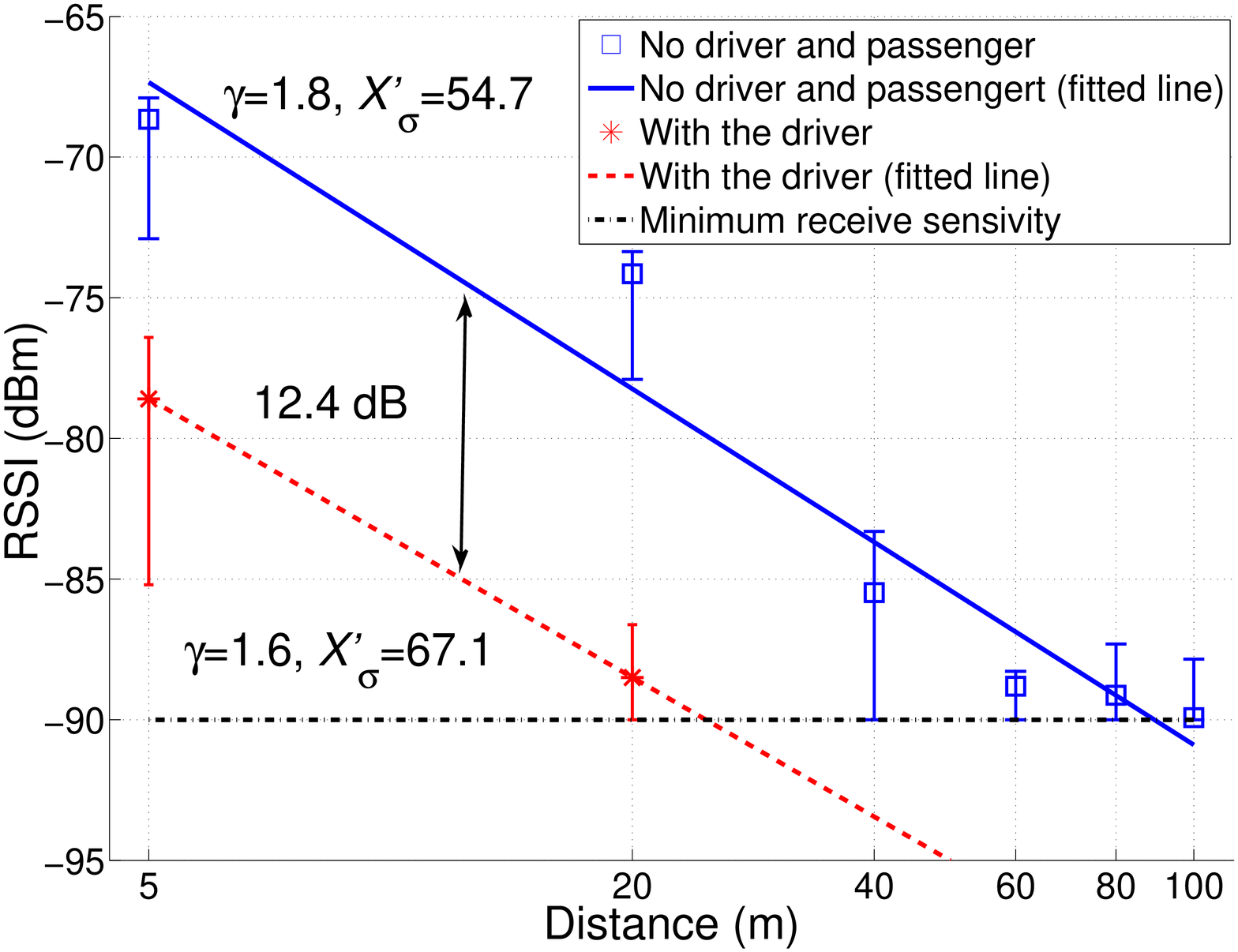}}
\caption{Comparison of RSSI with and without the driver}
\label{fig:RSSIdriver}
\end{figure}

Table~\ref{tab:Comparison1} shows the comparison of attenuation and path loss exponent values of tested antenna locations. One can observe that the path loss exponent values obtained from the fits are all close to free space path loss exponent (2). This is expected; with our experimental setting, besides the components in the scooters, there is no other obstacles that could result in additional signal attenuation and a larger path loss exponent. \michael{In addition, with the antennas used for our experiments, the two-ray ground reflection model is not a good approximation as the antenna gain in the direction of the ground-reflected ray for both the transmitter and the receiver is small due to the short T-R distance.} One can also observe that the location on the left mirror can potentially be the best location for mounting the antenna as the results suggest that it provides the least signal attenuation, with 10 - 20 dB less loss compared to some of the other locations, as it is the highest location on a scooter.

The propagation path when the antenna is located at the top of the headlight or the left of the headlight is significantly affected by the scooter frame. There is no line-of-sight between the transmitter and the receiver when the antenna is mounted on the top of headlight, as the path is blocked by the scooter frame. That is why the attenuation is the highest for this location. The scooter frame either fully or partially blocks the line-of-sight with every tested antenna location except the location at the left mirror. For example, the handle bar blocks the path when the antenna is mounted on the rear carrier and this \michael{on average} results in an additional attenuation of \michael{9.2 dB} compared to the left mirror location. 

\subsection{Human body shadowing: driver and passenger}
\label{subsec:rider_and_pax}

\begin{table}[t]
\centering
\caption{Additional mean attenuation caused by the body shadowing of the driver}
\resizebox{\textwidth}{!}{
\begin{tabularx}{1.5\textwidth}{@{} | >{\centering}p{0.3\textwidth} | >{\centering}X | >{\centering}X | >{\centering}X | >{\centering}X | X<{\centering} | @{}}
\hline
\textbf{Antenna Location}	&\textbf{Rear Carrier	}&\textbf{Top of the Headlight}	&\textbf{Top-Right of the License Plate}	&\textbf{Left Mirror} &\textbf{Left of the Headlight}\\ \hline
\hline
{$X'_\sigma$, \\ \textbf{scooter only}}&-43.4 dB&	-54.7 dB&	-34.6 dB&	-34.2 dB&	-49.1 dB\\

\hline
{$X'_\sigma$, \\ \textbf{scooter with the driver}}&	-53.9 dB&	-67.1 dB&	-44.2 dB&	-44.3 dB&	-58.5 dB\\

\hline \hline
{Additional attenuation caused by the driver (difference of the two $X'_\sigma$ values)}& 10.5 dB&	12.4 dB &	9.6 dB&	10.1 dB&	9.4 dB\\
\hline
\end{tabularx}
}
\label{tab:BodyAttenuation}
\end{table}

\begin{figure}[t]
\centering
\subfigure[Results comparing w/ and w/o driver and passenger]{\includegraphics[width=0.49\textwidth]{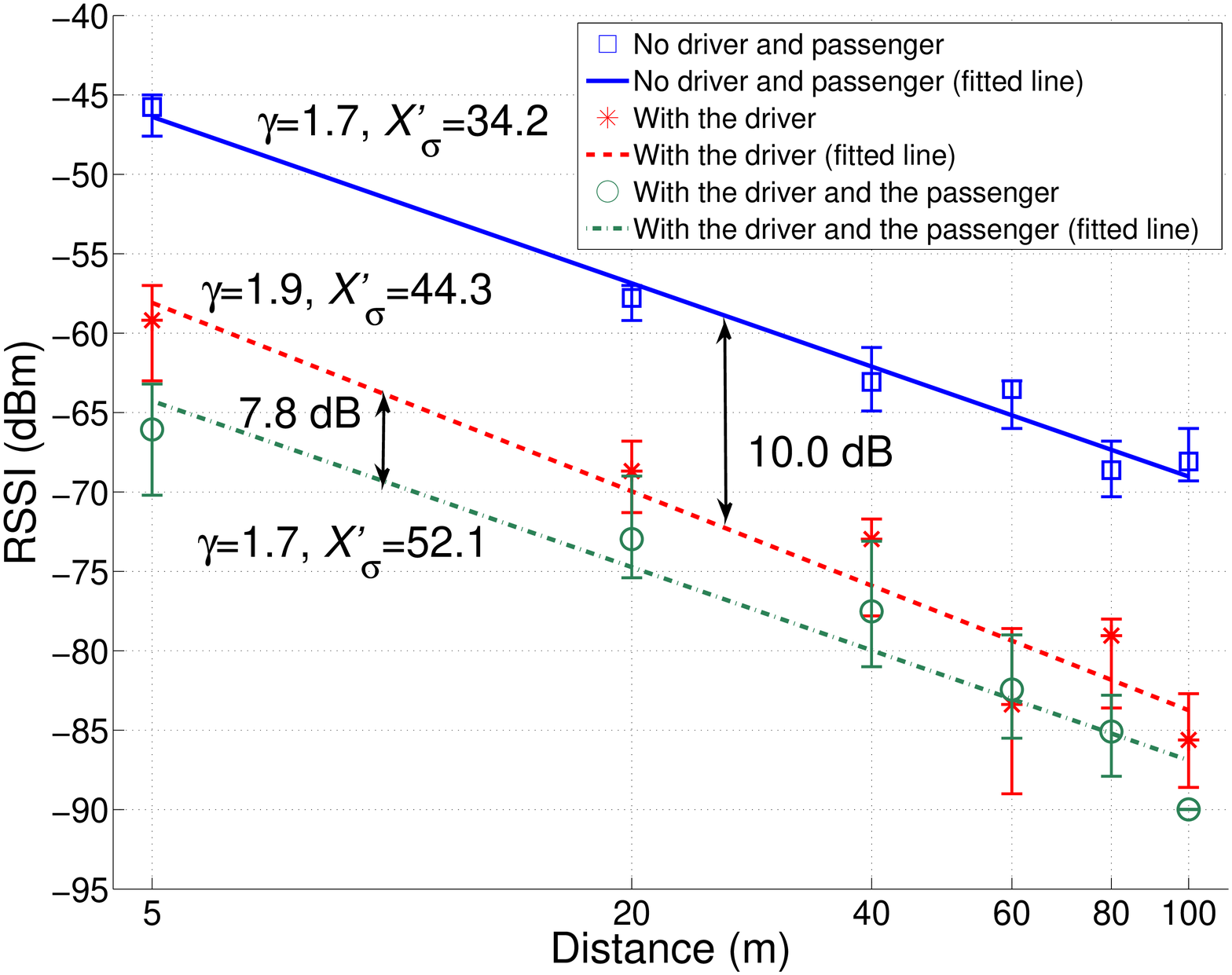}}
\subfigure[Results comparing different receiving nodes]{\includegraphics[width=0.49\textwidth]{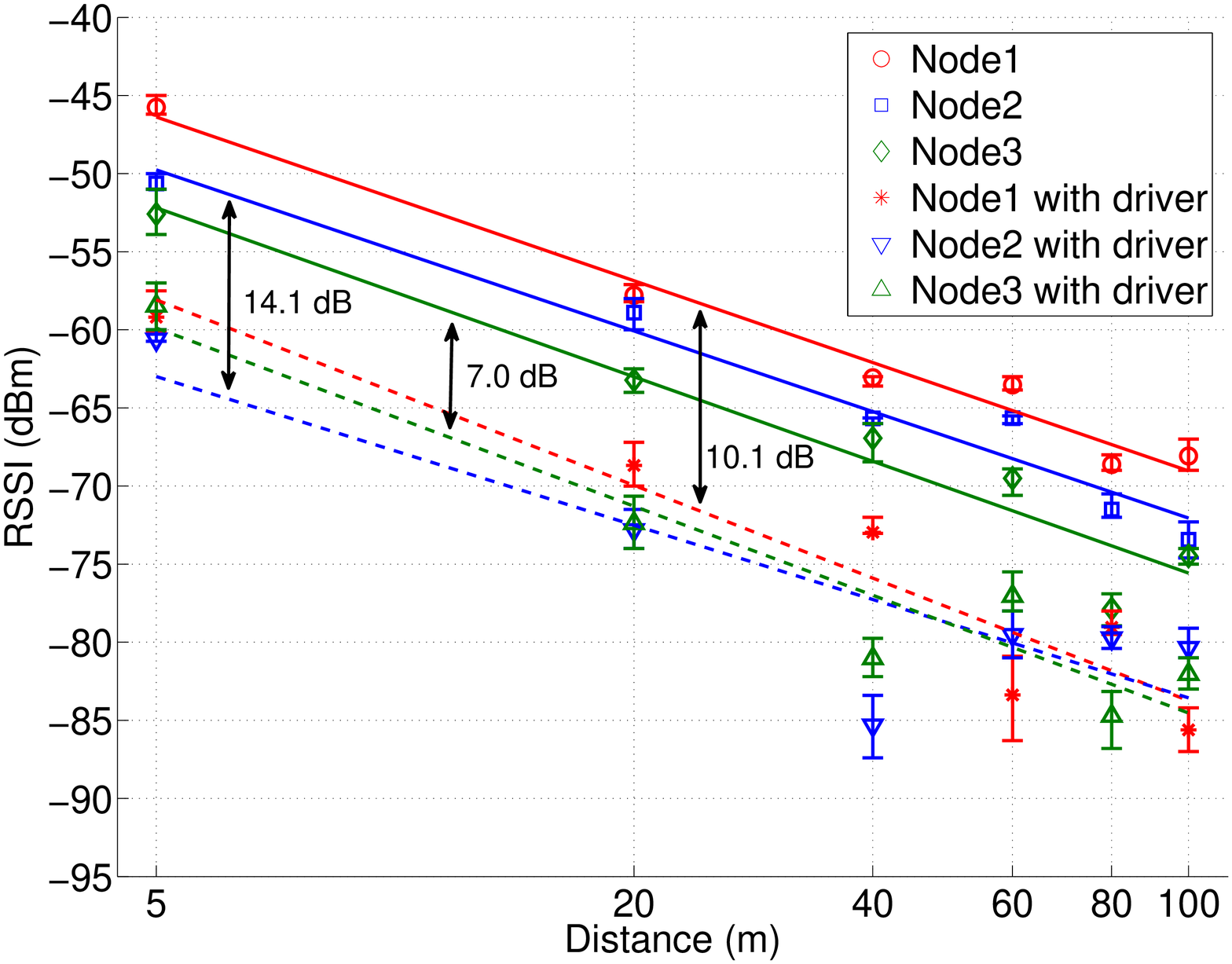}}
\caption{Comparison of RSSI with and without the driver at slightly different antenna locations. Antenna location: left mirror.}
\label{fig:RSSIRiderandPassenger}
\end{figure}

%

In the previous subsection, we investigated the amount of signal attenuation when different antenna locations are used on a scooter without a driver or a passenger.
However, in the real world, the driver and possibly a passenger ride on the scooter.
Therefore, in this subsection we repeat the same link measurement but with a driver and a passenger sitting on the scooter, in order to determine the impact of the driver and the passenger on the channel. 

In the first experiment, we asked a driver to sit on either the front scooter (receiver) or the rear scooter (transmitter). This depends on whether the location of the antenna is in front of the driver or behind the driver. When the antenna is in front of the driver, the driver is asked to sit on the front scooter (receiver), so that he would block some of the propagation paths between the transmitter and the receiver. In the other case, the driver is asked to sit on the rear scooter (transmitter), for the same reason.
The scooter driver is 173 cm in height, with shoulder width of 42 cm and torso thickness of 20 cm. 
Detailed dimensions of the scooter and the driver are shown in Figure~\ref{fig:DriverPassengerDimension}(a) and (b). 

\michael{To show that the human body is in the far field with respect to the transmitting antenna, we calculate the Fraunhofer distance, $d_f=\frac{2 D_{ant}^2}{\lambda}$ where $D_{ant}$ is the largest dimension of the antenna and $\lambda$ is the wavelength~\cite{Rappaport2001}. The Fraunhofer distance separates the near field and the far field, and is approximately 50.6 cm with our setting. Even with the antenna location that is the closest to the human body (left mirror), the distance between the antenna and the human body is approximately 45.5 cm and the human body locates almost entirely in the far-field.}

Figure~\ref{fig:RSSIdriver} and Table~\ref{tab:BodyAttenuation} compare the RSSI values with and without the driver.
As shown in Figure~\ref{fig:RSSIdriver}(a), the average RSSI values obtained with the antennas placed next to the left mirror show an average difference of 13.7 dB comparing the cases with and without the driver. The distance between the best fitting linear functions ($X_\sigma'$) is 10.1 dB. The difference in these two values is caused by small-scale fading, which is more prominent when the T-R distance is large. Similar results can be observed for three of the other locations, the rear carrier, top-right of the license plate, and left of the headlight, with the differences of attenuation between the cases with and without the driver ranging from 9.4 dB to 10.5 dB. 
A more interesting case is shown in Figure~\ref{fig:RSSIdriver}(b), presenting the results obtained when the antenna is placed at the top of the headlight. \michael{With this antenna location, the signal attenuation is larger and the received signal strength drops below the received sensitivity threshold (-90 dBm) at larger T-R distances.}
Thus, only measurements with T-R distances at 5 and 20 meter are used for linear function fitting. The results show that the difference of signal attenuation with and without the driver is 12.4 dB and is still in line with the values observed from other antenna locations. The numbers obtained from this experiment are very close to those reported in~\cite{HumanTorso}.

We carried out a second experiment to investigate how much additional attenuation a passenger would introduce, in additional to that of the driver. The dimensions of the scooter, the driver, and the passenger is shown in Figure~\ref{fig:DriverPassengerDimension}(c) and (d). This experiment is only performed with the antenna placed at the left mirror, the best performing location with the least signal attenuation. The passenger and the driver are asked to sit on the front scooter (receiver). The passenger is 175 cm in height, with shoulder width of 42 cm and torso thickness of 20 cm. One can observe that the passenger extends both the height and the width of the front scooter. 

Figure~\ref{fig:RSSIRiderandPassenger}(a) compares the RSSI of three different settings, without the driver, with the driver, and with both the driver and the passenger. Compared to the setting with only the scooter, there is an additional 17.9 dB attenuation with both the driver and the passenger. Compared to the result with only the driver sitting on the scooter, approximately an additional 5 to 8 dB of attenuation is observed. 
\mate{The reason why the passenger does not create as much attenuation as the driver is twofold: 1) the driver's body already blocks the line-of-sight path; and 2) the additional angle occupied by the passenger is minimal, because the driver already covers most of the same angle. Therefore, the impact of the passenger is mainly on the transmission propagation path going through the bodies, which is only one of the propagation mechanisms: reflected, diffracted, and scattered paths around the bodies are for the most part not affected by the additional (passenger) body.}

\michael{Finally, to show that the difference in attenuation in the scenarios with and without the driver is not caused by small-scale fading, we compare the results obtained from 3 different receiving nodes in two experiments. Node 1 is used in the first experiment, while node 2 and node 3, placed approximately 6 cm\footnote{Approximately half of the wavelength for 2.4 GHz signal.} apart from each other, are used in the second experiment. The second experiment is carried out on a different day and thus the location of the scooters and the nodes are slightly different. Figure~\ref{fig:RSSIRiderandPassenger}(b) shows the results from the experiment. One can observe that in results from all 3 nodes the difference in attenuation in the scenarios with and without the driver are all apparent, while the actual value varies with the setting.}

In summary, human body shadowing creates non-negligible strong signal attenuation that could have a mean value up to 12 dB with the driver, and up to 18 dB with both the driver and the passenger. This is observed for all tested antenna locations and thus we believe that any scooter-to-X communication system needs to take this into consideration to avoid performance estimation that could be too optimistic otherwise. \michael{We also confirm the results from the previous subsection that the left mirror is also the optimal antenna location with the least attenuation when the driver is present.
It is also worth noting that, while theoretically the amount of additional attenuation decreases as the T-R distance increases, as human bodies blocks less propagation paths with larger T-R distance, in our experiments with T-R distance up to 100 meters we observe approximately fixed amount of additional attenuation. Further characterization is needed to identify the offset due to the change of T-R distance and is left as future work.}

\subsection{Effects of driving lane}
\label{subsect:drivinglane}

%
%

\begin{figure}[t]
\centering
\subfigure[With line-of-sight]{
\includegraphics[width=0.25\textwidth]{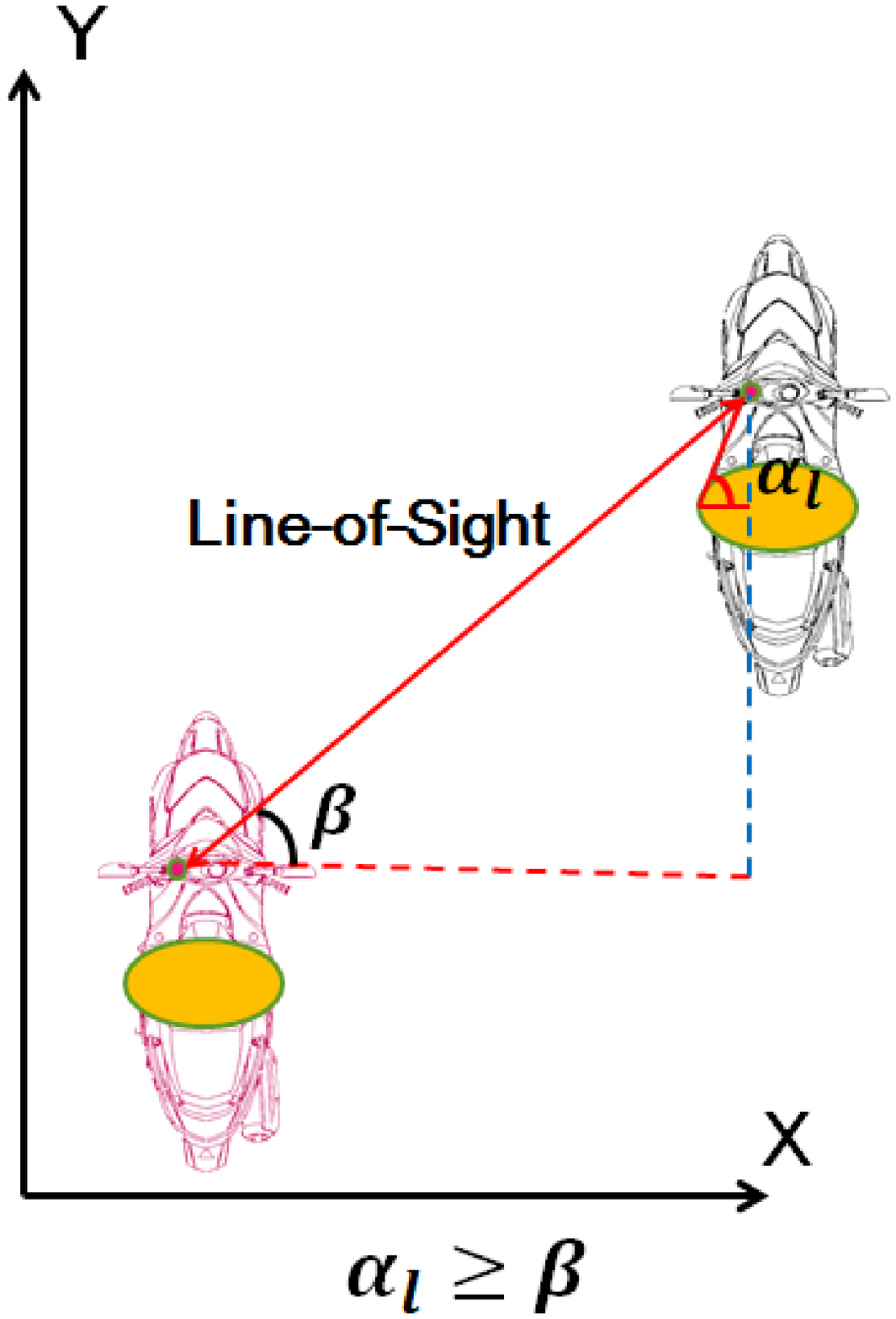}
\label{LaneEffectEnva}
}
\subfigure[Without line-of-sight]{
\includegraphics[width=0.25\textwidth]{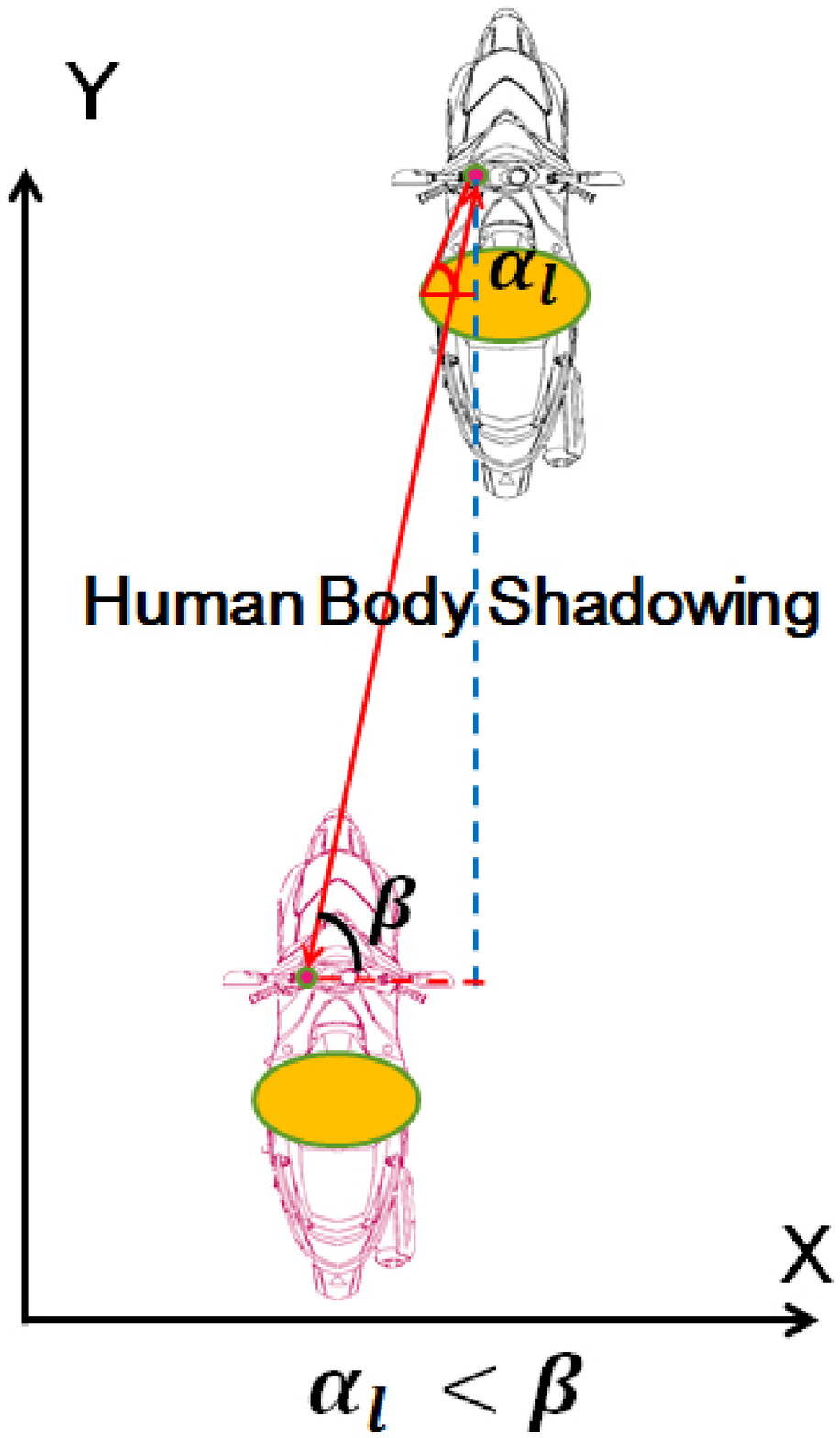}
\label{LaneEffectEnvb}
}
\subfigure[Used notations]{
\includegraphics[width=0.3\textwidth]{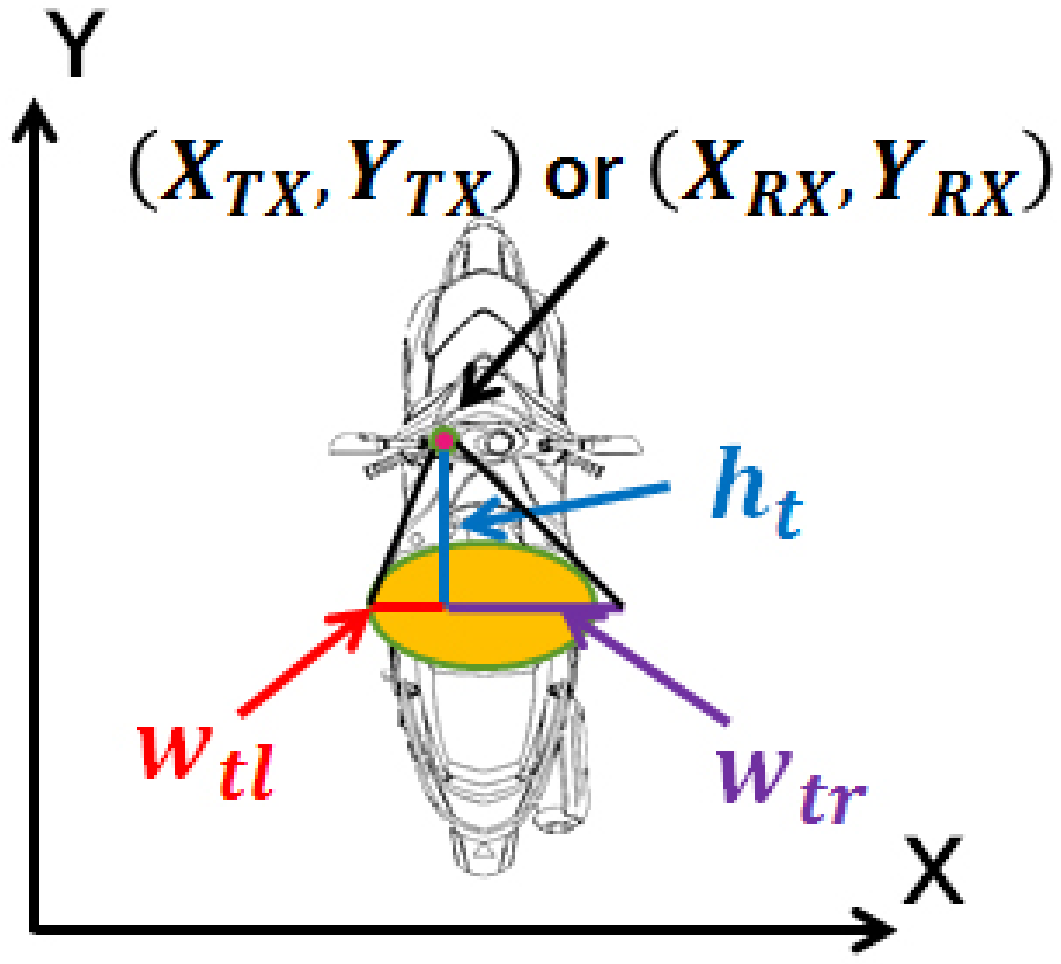}}
\caption{Relative positions of scooters under different propagation conditions and used notations}
\label{fig:LaneEffectEnv}
\end{figure}

\begin{figure}[t]
\centering
\includegraphics[width=0.8\textwidth]{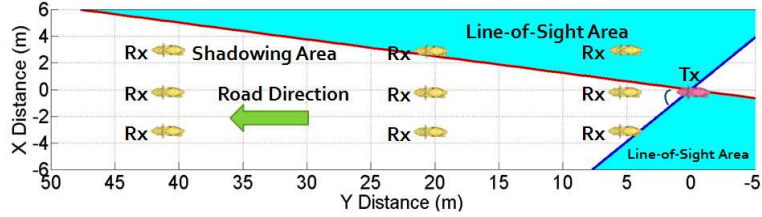}
\caption{Line-of-sight areas, when the receiver is in different positions with respect to the transmitter}
\label{fig:LineOfSightPrediction}
\end{figure}

\begin{figure}[t]
\centering
\includegraphics[width=0.5\textwidth]{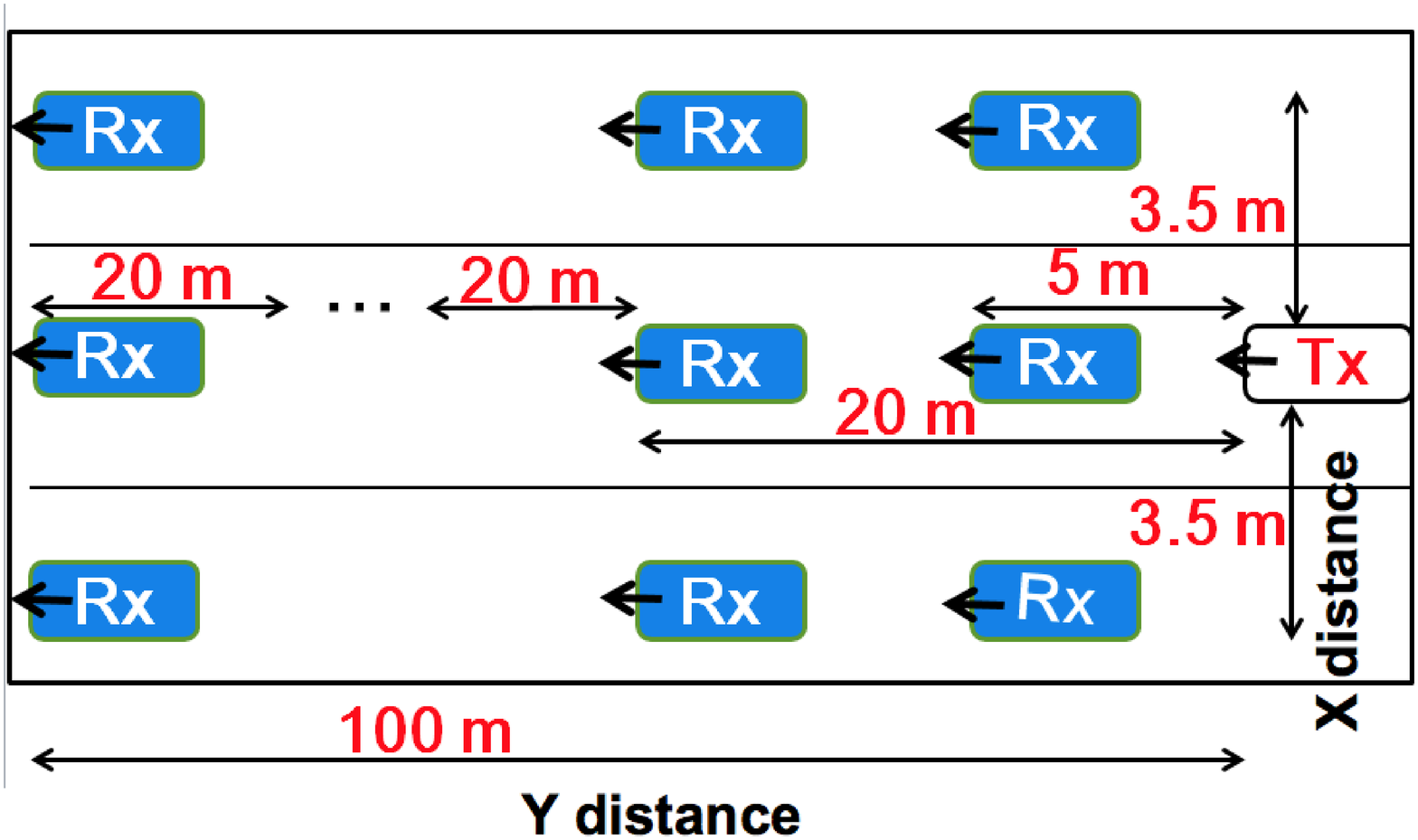}
\caption{The experimental setup}
\label{fig:LaneEffectExpSetup}
\end{figure}

\begin{figure}[t]
\centering
\includegraphics[width=0.6\textwidth]{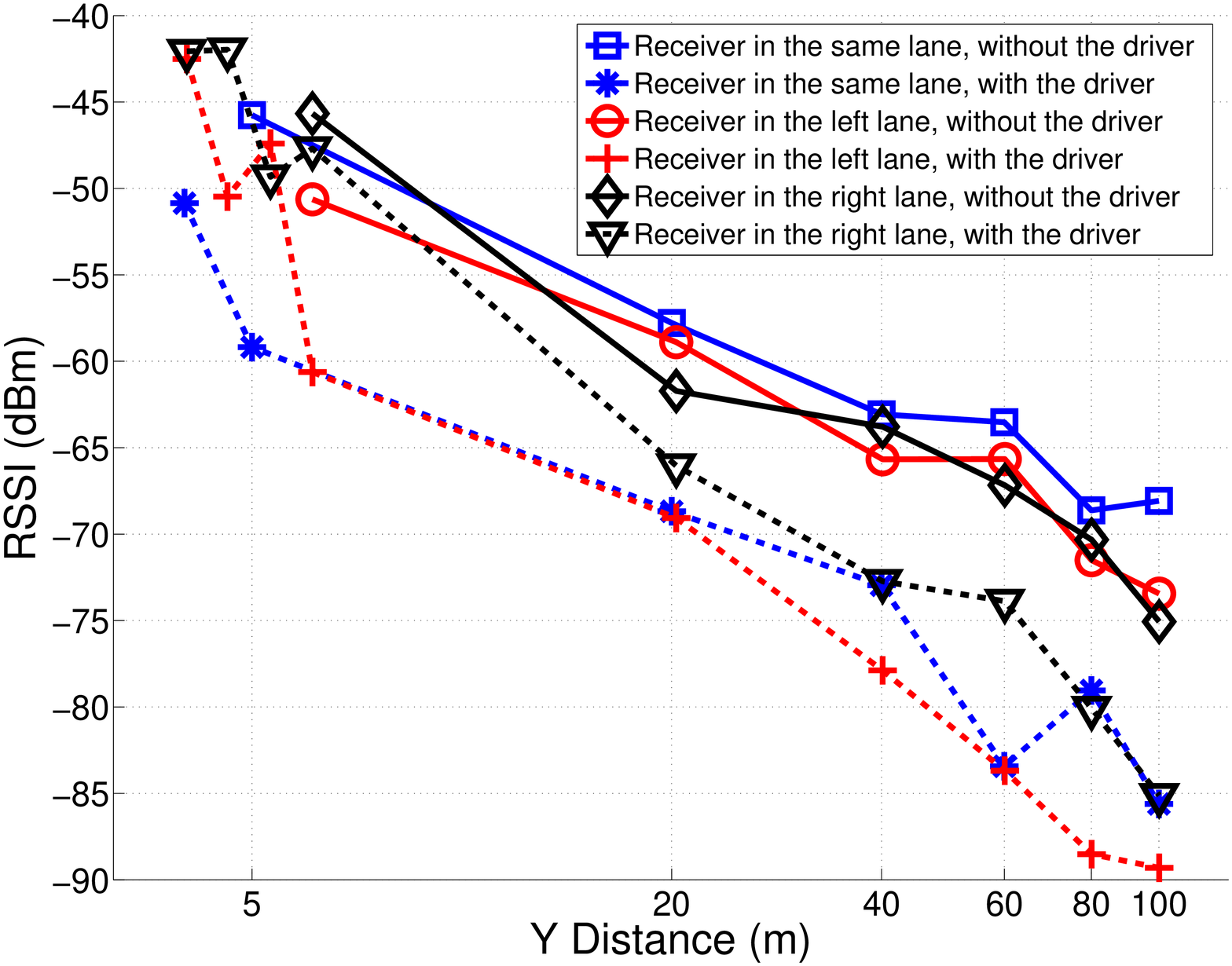}
\caption{Comparison of RSSI when the receiver is in different lanes}
\label{fig:LaneEffectCompare}
\end{figure}

In the real world, there can be multiple lanes on the road. In this subsection, we design a model to determine whether human body shadowing is in effect according to the relative positions of the transmitter and the receiver, which is usually determined by their driving lanes. Based on the measurement results, in the following analysis, we assume that the antenna is mounted on the left mirror of the scooter. 

We first consider the case that the transmitting scooter stays in \textit{front} of and on the \textit{right} side of the receiving scooter. Let $\alpha_l$ denote the angle between the central horizontal axis of the human body and the line connecting the antenna location and the left most position of the human body. Let $\beta$ denote the angle between the horizontal line going through the receiver's antenna location and the line connecting the receiving and the transmitting antennas. As shown in Figure~\ref{fig:LaneEffectEnv}(a)(b), it is obvious that there is line-of-sight between the two antennas \textit{if and only if the inequality holds:}
\begin{equation}
\alpha_l \geq \beta \qquad \textrm{.}
\label{eq:los1}
\end{equation}

Let $P_{\operatorname{TX}}=(X_{\operatorname{TX}},Y_{\operatorname{TX}})$ and $P_{\operatorname{RX}}=(X_{\operatorname{RX}},Y_{\operatorname{RX}})$ denote the coordinates of the transmitting and the receiving antennas, respectively. Let $h_t$ denote the distance between the antenna location and the central horizontal axis of the human body and $w_{tl}$ denote the distance between the left most position of the human body and the projected position of the antenna location on the central horizontal axis of the human body (see Figure~\ref{fig:LaneEffectEnv}(c)). Then, eq. (\ref{eq:los1}) can be rewritten as
\begin{equation}
\frac{h_t}{w_{tl}} \geq \frac{ | Y_{\operatorname{TX}} - Y_{\operatorname{RX}} | } { | X_{\operatorname{TX}} - X_{\operatorname{RX}} | } \qquad \textrm{.}
\label{eq:los2}
\end{equation}

Now, we will consider the case that the transmitting scooter stays in \textit{front} of and on the \textit{left} side of the receiving scooter. Let $w_{tr}$ denote the distance between the right most position of the human body and the projected position of the antenna location on the central horizontal axis of the human body (see Figure~\ref{fig:LaneEffectEnv}(c)). Then, similar to (\ref{eq:los2}), we can establish an inequality
\begin{equation}
\frac{h_t}{w_{tr}} \geq \frac{ | Y_{\operatorname{TX}} - Y_{\operatorname{RX}} | } { | X_{\operatorname{TX}} - X_{\operatorname{RX}} | } \qquad \textrm{.}
\label{eq:los3}
\end{equation}

Next, we will consider the two remaining cases where the transmitting scooter stays \textit{behind} the receiving scooter. Since, when an object blocks the line-of-sight path between the two antennas, it is true for both transmission directions, and the RX and TX related terms in eq. (\ref{eq:los2}) and eq. (\ref{eq:los3}) are symmetric, we can reuse these two inequalities. 

Combining all 4 cases, we can establish the following inequalities to test if two scooters have line-of-sight and are affected by human body shadowing:
\begin{equation}
\left \{
\begin{array}{l r}
\frac{h_t}{w_{tr}} \geq \frac{ | Y_{\operatorname{TX}} - Y_{\operatorname{RX}} | } { | X_{\operatorname{TX}} - X_{\operatorname{RX}} | } , & \textrm{if } X_{\operatorname{TX}} < X_{\operatorname{RX}} \textrm{;}\\
\frac{h_t}{w_{tl}} \geq \frac{ | Y_{\operatorname{TX}} - Y_{\operatorname{RX}} | } { | X_{\operatorname{TX}} - X_{\operatorname{RX}} | } , & \textrm{otherwise.}
\end{array}
\right .
\label{eq:los_all}
\end{equation}

Using eq. (\ref{eq:los_all}), in Figure~\ref{fig:LineOfSightPrediction} we illustrate the areas where the two antennas have line-of-sight, i.e., are not affected by human body shadowing, when the receiver is in different locations with respect to the transmitter; the figure is generated using the dimensions shown in Figure~\ref{fig:DriverPassengerDimension}. 

In order to validate the proposed model, we carried out another series of link measurements, where the receiver is placed at locations shown in Figure~\ref{fig:LaneEffectExpSetup}. 
\textblue{As shown in the figure, we define ``X distance'' as the lateral distance between the two scooters in the direction perpendicular to the road and ``Y distance'' as the longitudinal distance between the scooters in the direction of the road.} 
The X distance between the transmitter and the receiver is either 0 or 3.5 meters, emulating both the cases that they stay in the same or different lanes. In this series of measurements, the experiments are carried out both with and without the driver on the scooter, in order to observe whether human body shadowing affects the channel attenuation. Using the result shown in Figure~\ref{fig:LineOfSightPrediction}, we expect that, when the receiver stays in the left lane and the right lane with respect to the transmitter, human body shadowing starts to become prominent when the Y distance between the two scooters is larger than 3.9 meters and 24 meters, respectively. 

Figure~\ref{fig:LaneEffectCompare} shows the measurement results, which are in line with our expectation. Comparing the RSSI with and without the driver when the receiver is on the \textit{left} of the  transmitter, the difference becomes apparent after the Y distance between the two scooters is larger than 5 meters. On the other hand, comparing the RSSI with and without the driver when the receiver is on the \textit{right} of the transmitter, the difference becomes apparent until the Y distance between the two scooters is larger than 20 meters. These results confirm the validity of our proposed model for determining whether human body shadowing is in effect. 

The equations above (eq.~\ref{eq:los1}-\ref{eq:los_all}) determine whether or not a scooter-to-X link is obstructed by a body of a scooter driver or passenger. These equations should be utilized in concert with models determining existence/non-existence of other types of objects (such as obstructing vehicles, buildings, etc.) in the line-of-sight path of the link, in order to determine whether the link is shadowed or not. For this purpose, in Section 4.3, we perform simulations where we combine the human body shadowing calculations with the model proposed by Boban et al.~\cite{boban14TVT,boban11}, which determines shadowing effect of other vehicles, as well as buildings and foliage.

\subsection{Verification: driving scenario}

\begin{figure*}
\centering
\subfigure[Receiver in the front]{\includegraphics[width=0.48\textwidth]{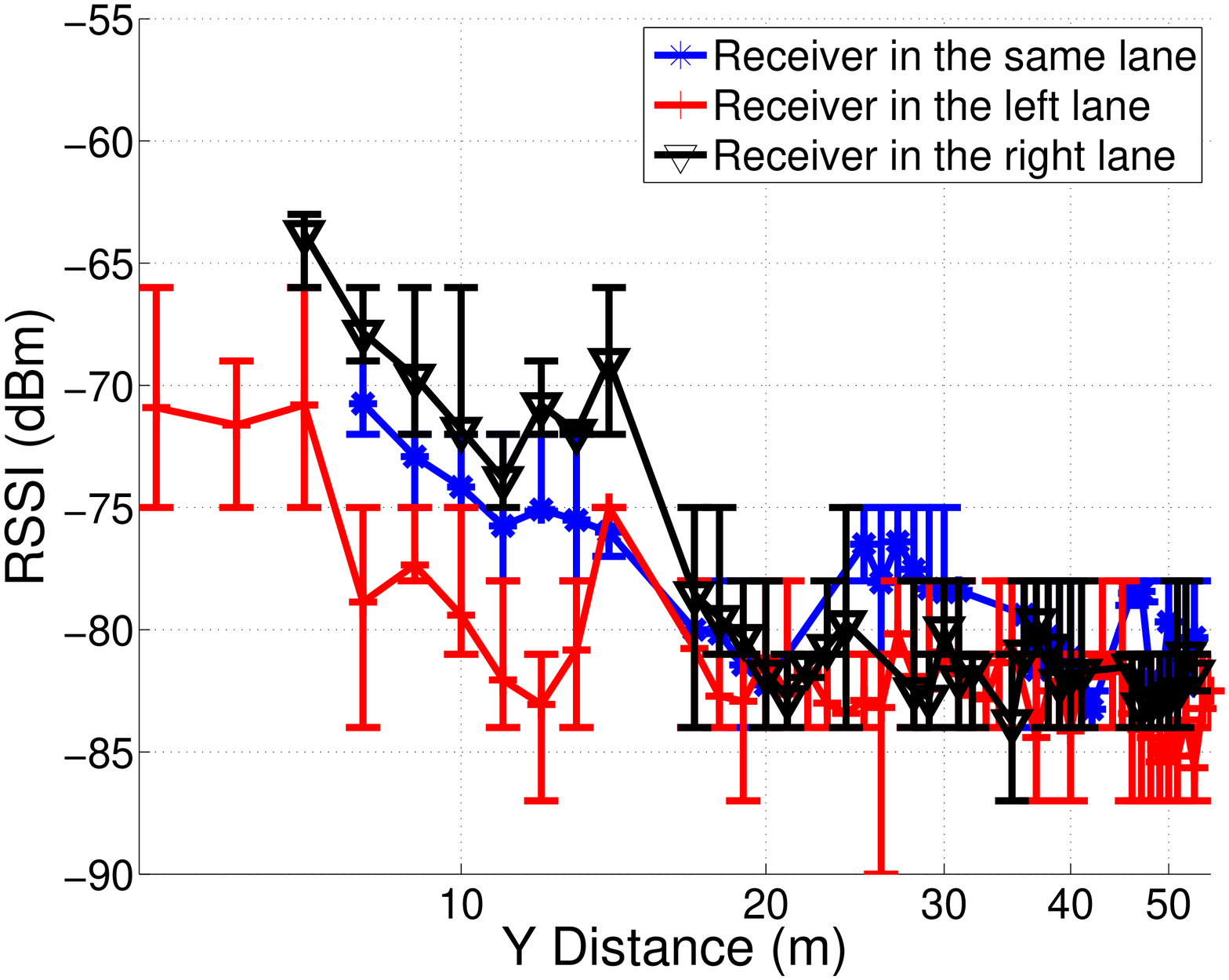}}
\subfigure[Transmitter in the front]{\includegraphics[width=0.48\textwidth]{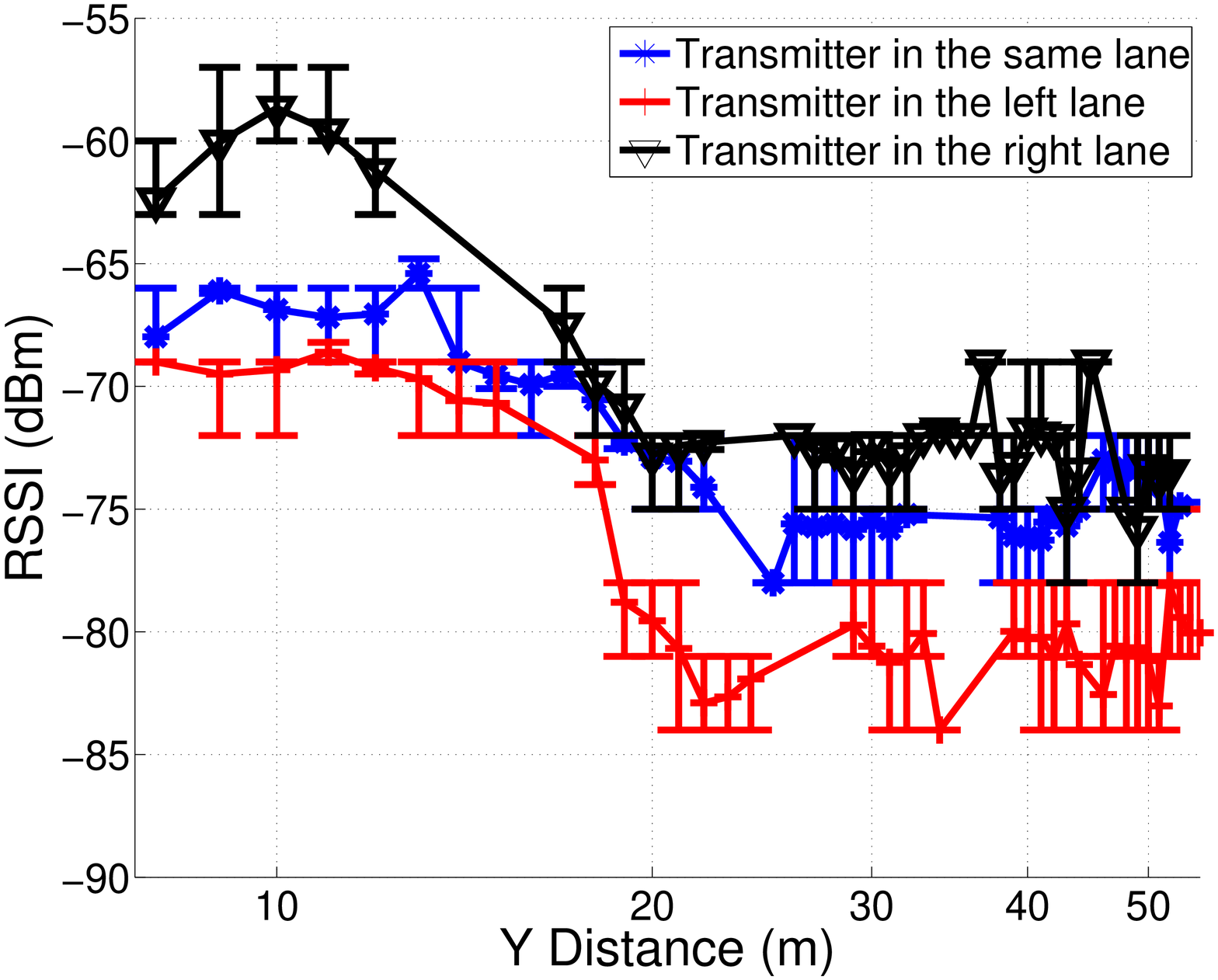}}
\caption{\textred{Comparison of RSSI when} \textblue{the back scooter is in the middle lane while the front scooter} \textred{is in different lanes in moving scenarios. The markers represent the mean RSSI while the bounds of the error bars represent 25\% and 75\% percentiles of RSSI.}}
\label{fig:mobile}
\end{figure*}


\textred{To verify that human body shadowing as well as the effects of driving lane take place even when the scooters are in motion, a small set of measurements are performed at the same empty and closed road segment, with both scooters traveling at normal speed. We mounted a 2-dimensional LADAR, SICK LMS-291~\cite{SICK}, at the back of the front scooter, to accurately measure the distance to the back scooter. The LADAR can measure up to 50 meters, given the driving scenarios and the reflectivity of the scooters, while producing extremely accurate distance measurements with centimeter-level error.} \textblue{The two drivers were instructed to attempt to maintain the distance at 10, 20, 30, 40, or 50 meters, while they traveled at approximately 20 km/h.
The back scooter was kept in the middle lane while the front scooter traveled in the left, middle, and right lanes in different runs.} 
 \textred{The scooters traveled for approximately 150 meters in each run. Two batches of measurements are carried out, with the front and the back scooter being the transmitter and the other being the receiver, respectively. Data from a total of 26 runs (13 runs per batch) are collected. Figure~\ref{fig:mobile} shows the results. Packets received at different distances are first split into equal-sized bins of 1 meter. Then, the mean and the 25\% and 75\% percentiles of RSSI are calculated for each bin.}

\textblue{Figure~\ref{fig:mobile} shows that the effect of driving lane is still quite prominent, despite the effect of small-scale fading resulting from the motion of the scooters, which} \textred{creates fluctuations spatially at different distances and also at the same distance over time. The trend in the figures is identical to that in Figure~\ref{fig:LaneEffectCompare}. When the front vehicle is on the left, human body shadowing does not take place until the Y distance is more than approximately 20 meter, where the black curves start to merge with the blue and the red ones. The red curves constantly have the smallest RSSI, as when the front vehicle is on the left, human body shadowing is always in effect for the range of distances we carried out the measurement. The blue curves usually sit between the black curves and the red curves.} \textblue{This is because, while the two scooters drive in the same lane, the front scooter can still be slightly to the left or to the right of the back scooter in the lateral direction as they travel, and the mean RSSI produces an average of the two cases.} \textred{In summary, we have verified that human body shadowing as well as the effect of driving lane can be clearly observed when the scooters are in motion.}

\section{Discussion}
\label{sec:discussion}
\subsection{Single Antenna Placement}
\textbf{The height of the antenna.} Many previous studies suggest that it is beneficial to install the antenna of a vehicular communication system at the \textit{highest} location of the car, i.e., the rooftop of a car, a van, or a bus~\cite{SECON07,obstructions10,boban11}, so that the metal vehicle body of other vehicles has a lower probability to block the line-of-sight path between the transmitting and the receiving antennas and to create additional channel attenuation. 
Our measurement results provide an additional reason to strengthen this design guideline; if the antenna on a vehicle is installed at a height that is lower than the height of the human body sitting on a scooter, when the propagation path penetrates the location of a scooter, there could be significant additional signal attenuation ranging from 12 to 18 dB. The typical height of a car ranges from 1.3 to 1.5 meters, which implies that the top of a car is probably the only feasible location that can be clear of the human body sitting on a scooter and avoid this additional attenuation (see Figure~\ref{fig:DriverPassengerDimension}(a)(c)).

The rationale also applies to scooters, either as the transmitter or the receiver. However, the same convention cannot be used since the highest location is the top of the driver's or the passenger's helmet; it is obvious that it is not \michael{very practical} to mount the antenna on the top of the helmet due to several reasons: inability to connect the system on the scooter and the system on the helmet (which can be the antenna only, or the entire communication system with the antenna) with a cable, significant channel variation caused by frequent movements of the helmet, etc.
Our measurement results suggest that the best antenna location on a scooter is the next to the rear-view mirror, which agrees with the guideline as it is the highest location on the scooter itself. However, along with all possible antenna locations other than the top of the helmet, they do not have sufficient height to avoid human body shadowing.
\begin{table}[t]
\centering
\caption{Cases in which Human Body Shadowing is in Effect}
\begin{tabular}{|c||c|c|}
\hline
                 & \multicolumn{2}{c|}{\textbf{The leading vehicle is}} \\ \hline
\textbf{The trailing vehicle is}    & \textbf{a scooter} & \textbf{a car} \\ \hline \hline
\textbf{a scooter} & \textblue{body shadowing possible} & no shadowing \\ \hline
\textbf{a car}     & \textblue{body shadowing possible} & no shadowing \\ \hline

\end{tabular}
\label{tab:shadowing}
\end{table}

\textbf{The horizontal location of the antenna.} The shadowing area is the area where a vehicle experiences human body shadowing with respect to the host vehicle (which is shown as the shadowing area in Figure~\ref{fig:LineOfSightPrediction}) when establishing a communication link with the host vehicle. Whether the shadowing effect takes place depends on the type of communicating vehicles and their relative position.

Table~\ref{tab:shadowing} summarizes whether the shadowing effect takes place, considering all cases when the vehicles travel in the same direction. For example, when a car is located in front of a scooter and the two vehicles establish a link, there is no human body shadowing since the antenna is in front of the human body on the scooter. However, there could be human body shadowing when the scooter is in front of the car, since the human body could be located between the antennas of the two vehicles in some cases. The results for the other cases can be obtained in a similar manner. 

The central angle of the shadowing area, when the shadowing effect takes place, is dictated by eq.~(\ref{eq:los3}) and given by
\begin{equation}
\theta_{s}=\tan^{-1} \left( \frac{h_t}{w_{tl}} \right) + \tan^{-1} \left( \frac{h_t}{w_{tr}} \right) \qquad \textrm{.}
\end{equation}
With the dimensions of the scooters and the driver in our experiments, $\theta_s$ is approximately 45 degrees.

To mitigate the impact created by human body shadowing, one approach is to minimize $\theta_s$, either by increasing $w_{tl}$ and $w_{tr}$ or decreasing $h_t$. The former is infeasible, as the dimensions of the driver cannot be changed. The latter is limited by the dimensions of the scooter, and also hard to be satisfied. The other approach is to change the location of the antenna along the horizontal axis, changing the portion of $w_{tl}$ and $w_{tr}$ with respect to their sum. However, with this approach, $\theta_s$ would almost stay the same, while the shadowing area and the line-of-sight area would slightly rotate with respect to the antenna location.

One can observe from Figure~\ref{fig:LineOfSightPrediction} that the shadowing area almost points in the direction of the road. This means that \textit{when the T-R distance is large, human body shadowing is always in effect}. This also means that rotating the shadowing area slightly to the side could be helpful, at least for cases when the T-R distance is small. However, since the width of the scooter is approximately the same as the width of the human body, it severely limits the furthest the antenna can be moved to the side and the largest possible rotation angle. Nevertheless, we still recommend a design guideline whereby, in azimuth,the antenna should be located as far from the center of the scooter as possible.

\subsection{Dual Antenna Consideration}
\begin{figure}[t]
\centering
\subfigure[Dual front antennas]{\includegraphics[width=0.25\textwidth]{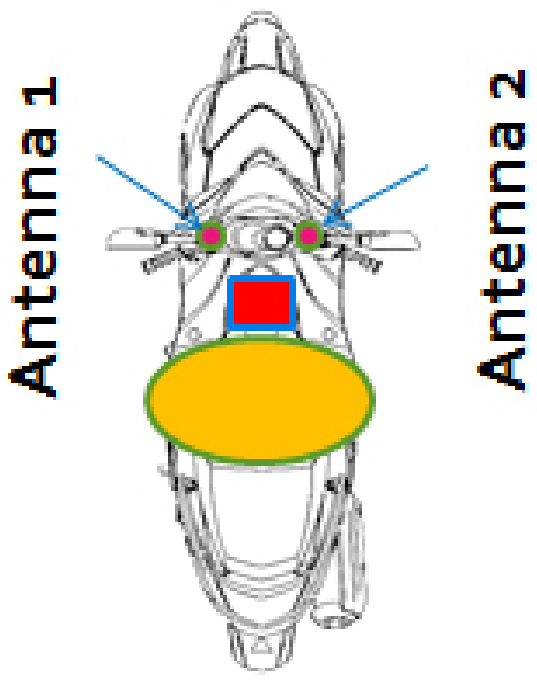}}
\subfigure[Front-and-back antennas]{\includegraphics[width=0.25\textwidth]{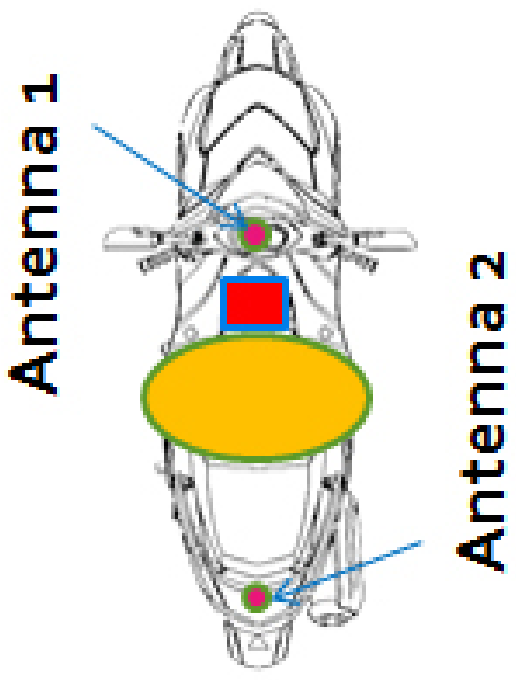}}
\caption{Two considered dual antenna configurations}
\label{fig:dual_antenna}
\end{figure}

The discussion in the previous subsection suggests that using a single antenna in the system cannot avoid human body shadowing in most cases. Another possible approach is to add a second antenna to the system, increasing the cost. The system can utilize techniques such as selective combining or maximal ratio combining~\cite{WirelessCommunications} to obtain a signal with larger received power; in the case that only one of the two antennas experience human body shadowing, the signal from the other one can be used and thus human body shadowing does not affect the receiving performance. In the following discussion, we consider two very simple dual antenna configurations, as shown in Figure~\ref{fig:dual_antenna}.

\textbf{The front-and-back antenna configuration} can completely eliminate human body shadowing when used in combination with selective combining. This is because, in any case, one of the two antennas will be able to provide a propagation path without going through the human body. However, additional RF cables need to be installed to connect the two antennas to the RF input/output of the system. These RF cables can be up to 4 meters in length\footnote{We actually installed these RF cable in one of our scooters and measured the length.}, creating an additional signal attenuation of almost 5 dB, which is already a significant portion of the attenuation that could be caused by human body shadowing. In addition, the cost of the cable is in the range of \michael{tens of} U.S. dollars. While this configuration can avoid human body shadowing, non-negligible cable loss and higher system cost, caused by the extra antenna, added circuitry, and RF cabling, renders it \michael{less attractive}.

\begin{figure*}[t]
\centering
\includegraphics[width=0.8\textwidth]{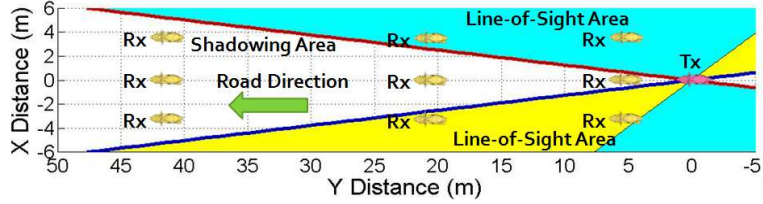}
\caption{Line-of-sight areas, dual front antenna configuration}
\label{fig:los_dual_front}
\end{figure*}

\begin{figure*}[t]
\centering
\subfigure[Single antenna, left mirror]{\includegraphics[width=0.49\textwidth]{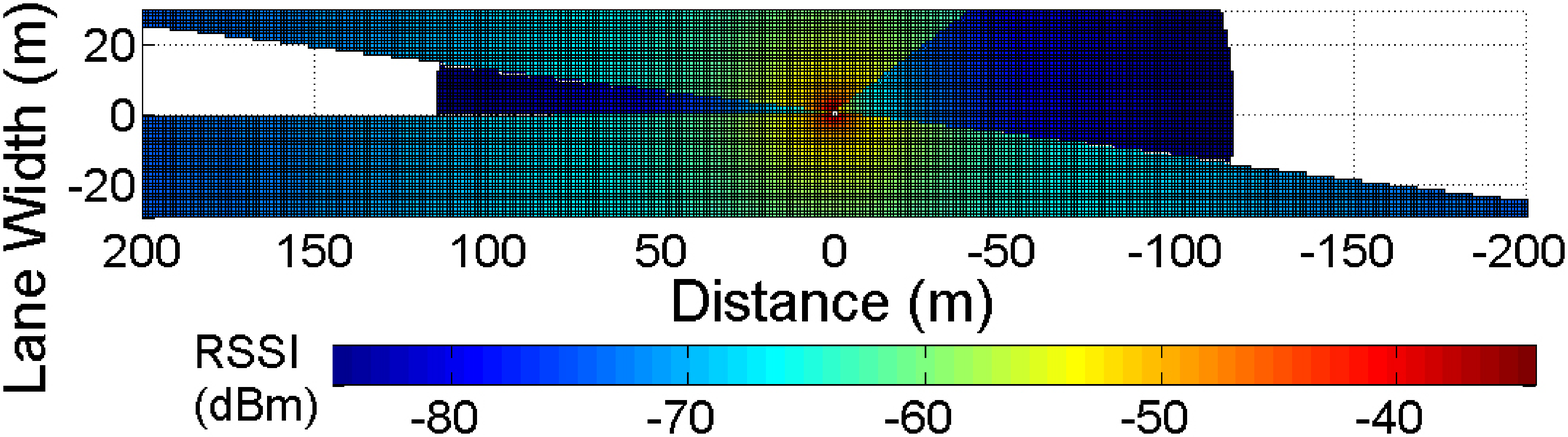}}
\subfigure[Dual front antenna]{\includegraphics[width=0.49\textwidth]{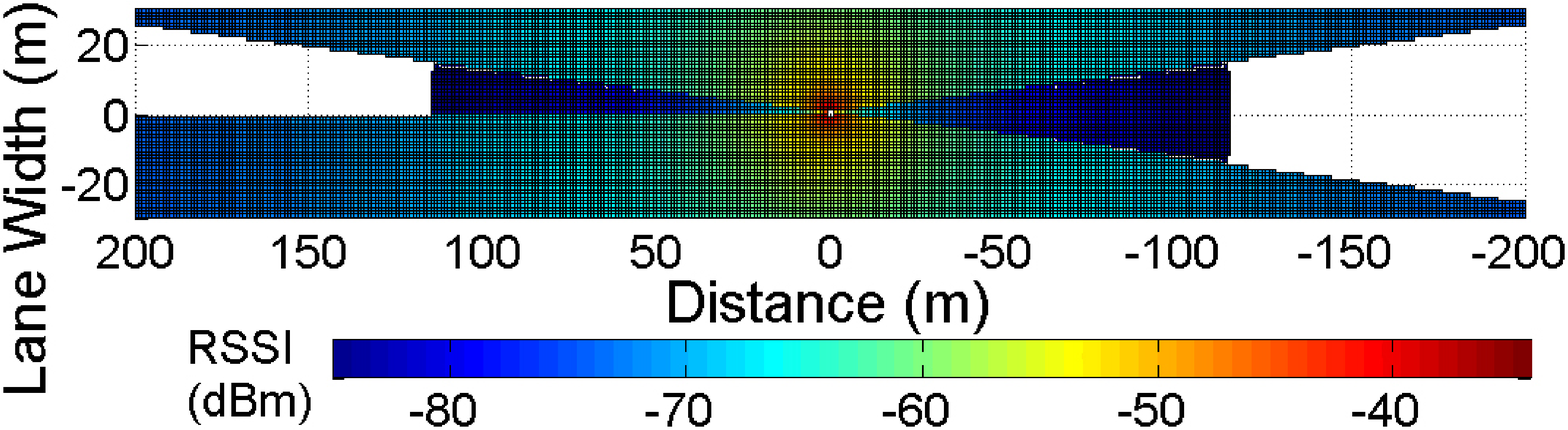}}
\caption{Comparison of the received signal strength at different locations with respect to the location of the transmitter, when using single antenna and dual front antennas. Note that typical width of a road is only 15-30 meters and that the coverage area when in line-of-sight condition goes beyond the boundary of the figure to 800 meters.}
\label{fig:coverage_comparison}
\end{figure*}

\textbf{The dual front antenna configuration} uses a second antenna on the opposite side of the original antenna mounted on the mirror, so that a portion of the shadowing area on the side can be eliminated, as shown in Figure~\ref{fig:los_dual_front}. While the central angle of the shadowing area reducing approximately from 45 degrees to 15 degrees seem a significant improvement, our investigation shows that the improvement of the actual coverage area in which a vehicle can receive the transmission is very limited. 

Figure~\ref{fig:coverage_comparison} compares the coverage area when using the single antenna configuration and the dual front antenna configuration. Based on our measurement results, the communication range is approximately 800 and 115 meters, without and with human body shadowing in effect, respectively. The receive sensitivity threshold is assumed to be -85 dBm. Observe in the figure that the area affected by human body shadowing spans along the direction of the road. The left side of the road is mostly not affected due to the opposite traveling direction. Our analysis indicates that the improvement of the coverage area when using the dual front antenna configuration is only 0.1\% compared to the single antenna configuration, which is negligible.

In summary, our investigation reveals that, although in theory a dual antenna configuration could utilize its diversity to mitigate the effect caused by human body shadowing, in practice it is hard to justify the additional system cost with the limited benefit. \textred{Under the cost constraints posed by the low sale price of the scooter, we} therefore believe that the solution to human body shadowing lies in the \textit{protocol design} instead of the \textit{physical layer design}. \textred{A more detailed discussion of the implications on the protocol design is presented in subsection~\ref{subsec:protocol}. }

\subsection{Impact on packet reception}
To understand the level of impact on packet reception in scooter-to-X communications, we implement the human body shadowing model reported in this paper in simulations. In addition to the body shadowing model, our simulations also take into account the time-varying V2V channel behavior by implementing the fading and obstruction caused by obstructing vehicles using the vehicles-as-obstacles model proposed by Boban et al.~\cite{boban11} and available in GEMV$^2$ simulator~\cite{boban14TVT}. In the simulations, each vehicle broadcasts its own location at 10 Hz with 100-byte packets. We implement the models in 
EstiNet~\cite{Estinet}, a commercial network simulator. \textblue{The simulations assume the use of 2.4 GHz IEEE 802.11b/g/n WiFi radio.} Simulation parameters can be found in Table~\ref{tab:Simulation1}.

\begin{table}[t]
\centering
\caption{Simulation Parameters}
\scriptsize
\begin{tabular}{c c}
\begin{tabularx}{0.6\textwidth}{@{} | >{\centering}X | X<{\centering} | @{}}
\hline
\multicolumn{2}{|c|}{\textbf{Main Parameters}} \\ \hline
\textbf{Parameter} & \textbf{Value} \\ \hline \hline
Simulation Time & 100 seconds\\
\hline
Number of Vehicles & 50\\
\hline
 Road Topology & Two connected square-shaped loop; each edge is 50 meters in length \\
\hline
Lane Width & 3 meters\\
\hline
Number of Lanes in Road & 3 lanes\\
\hline
Beacon Broadcast Frequency & 10 packets per second\\
\hline
Wireless Technology & IEEE 802.11b/g/n \\
 & (2.4 GHz WiFi) \\
\hline
Transmission Rate & Fixed, 11 Mbps \\ 
\hline
Packet Size & 100 bytes\\
\hline
Antenna Gain & 1 dBi\\
\hline
Carrier Sense Threshold & -87.57 dBm\\
\hline
Transmission Power & 15 dBm\\
\hline
\end{tabularx}
&
\begin{tabularx}{0.35\textwidth}{@{} | >{\centering}X | >{\centering}X | X<{\centering} | @{}}
\hline
\multicolumn{3}{|c|}{\textbf{Vehicle Profiles}} \\ \hline
\hline
 & \textbf{Scooter}  & \textbf{Car}\\
\hline \hline
Max Speed & 60 km/hr & 97 km/hr\\
\hline
Max Accel. & 8 m/s$^{2}$ & 7 m/s$^{2}$\\
\hline
Max Decel. & 4 m/s$^{2}$ & 5 m/s$^{2}$\\
\hline
Width & 1 m & 2 m\\
\hline
Length & 2 m & 5 m\\
\hline
Antenna Ht. & 113 cm & 143.5 cm\\
\hline
Obstacle Ht. & 166 cm\ & 143.5 cm\\
\hline
\end{tabularx}
\end{tabular}
\label{tab:Simulation1}
\end{table}


\begin{figure}[t]
\vspace{10pt}
\centering
\includegraphics[width=0.6\textwidth]{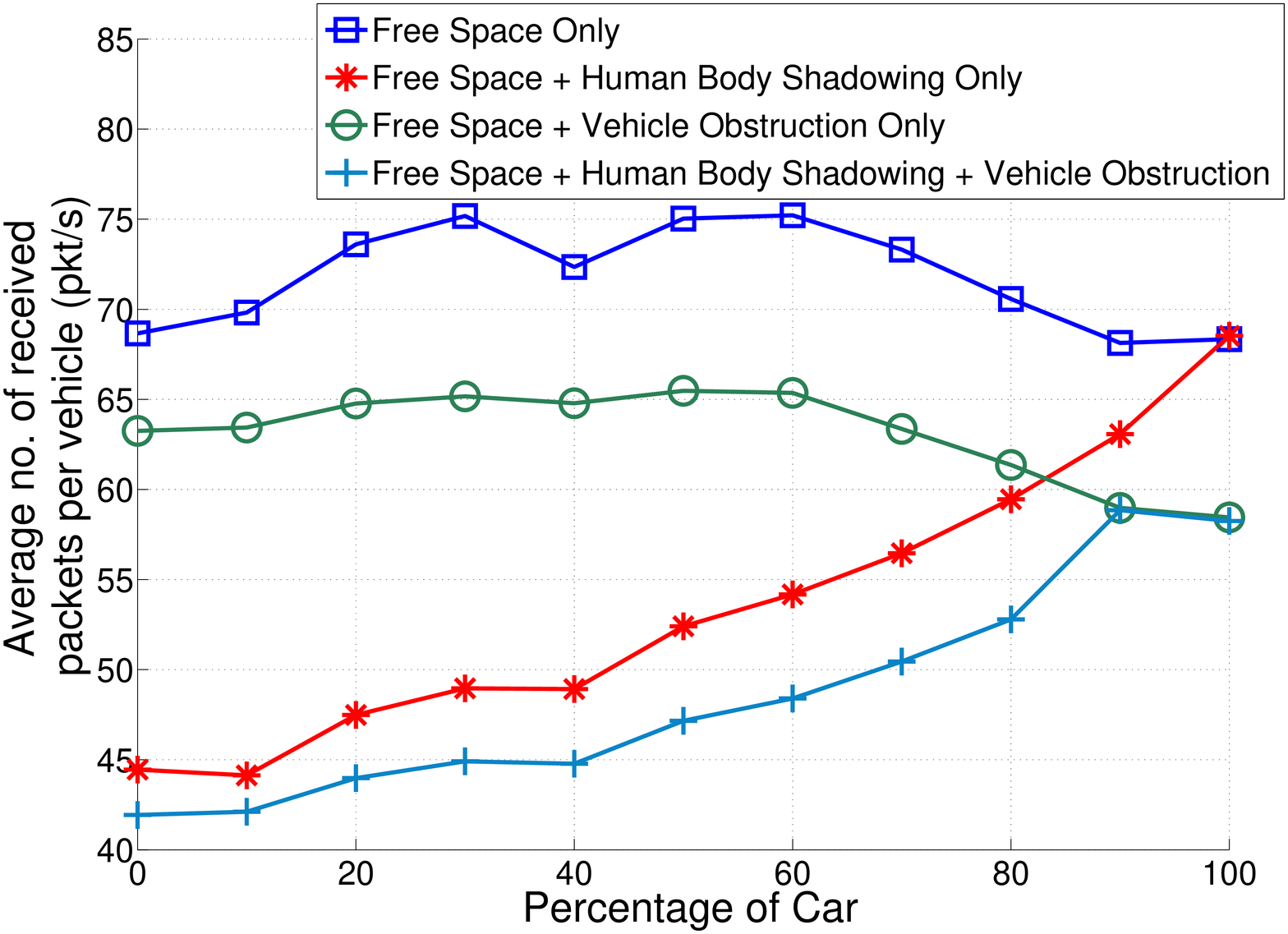}
\caption{Average number of received packets per vehicle}
\label{AvgInBroadcastIN}
\end{figure}

We carry out a number of simulations with different percentages of cars (while the others are scooters), ranging from 0\% to 100\%. For each percentage, we generate a set of mobility traces and the same traces are used in the simulations that incorporate different large-scale fading channel models, including (1) free-space only; (2) with additional attenuation caused by vehicle obstruction; (3) with additional attenuation caused by human body shadowing; and (4) with additional attenuation caused by both vehicle obstruction and human body shadowing. 

Figure~\ref{AvgInBroadcastIN} compares the average number of received packets per vehicle in different settings. One can observe that, in scenarios where there is a large percentage of scooters, there is drastic difference between the numbers of received packets for the case with human body shadowing and for the case without human body shadowing. For example, when there are 70\% scooters and 30\% cars\footnote{This is the ratio in Taiwan.}, the reduced number of received packets due to human body shadowing is as large as 40\%. The results also show that the impact of human body shadowing is generally larger than vehicle obstruction, when scooters contribute to more than 20\% of the vehicle population, with the impact of body shadowing being negligible only in scenarios with 10\% of scooters or less.

Simulation study or theoretical analysis are often used to conduct research to determine the feasibility or the performance of a certain application supported by vehicular communications or networks. These require accurate vehicle-to-vehicle channel models for the results to have high fidelity. Our results suggest that simulation or theoretical studies that involve scenarios with a large population of scooters should incorporate human body shadowing in their channel model; otherwise, the obtained simulation results are most likely going to be too optimistic.
\subsubsection{Implementing body shadowing model for scooter-2-X links in simulators}
Below we provide a simulation recipe for implementing scooter-2-X links in V2V simulation environments.
\begin{itemize}
\item Generate scooter mobility using microscopic traffic mobility simulators (e.g., SUMO~\cite{SUMO2012}, ideally with scooter mobility patterns~\cite{shih2011rule}).
\item Assign each scooter with driver and (optionally) passenger and their respective dimensions (available in Fig.~\ref{fig:DriverPassengerDimension})
\item Incorporate the LOS-existence calculations due to body shadowing from Sect.~\ref{sec:link_measurement}; geometric calculations can be carried out analogously to those explained in GEMV$^2$ simulator~\cite{boban14TVT}, including efficient geographic data manipulation with R-trees, quad-trees or similar data structure
\item For each scooter-to-X link, calculate the depth of transmission through body and the resulting signal attenuation 
\item Add the human body-based attenuation to the overall scooter-to-X link budget calculation, including path loss, fading, and shadowing by other objects (e.g., buildings and vehicles)
\end{itemize}

\subsection{Implications on protocol design}
\label{subsec:protocol}
An additional contribution of this study is that it can help to increase the reliability of scooter-to-X communications by providing a more realistic environment for protocol design and evaluation. 
\textred{We discuss the implications} \textblue{on} \textred{the protocol design with 3 different communication patterns as follows.}
\textred{
\begin{itemize}
\item \textbf{Direct transmission over a single link:} When the location of the intended receiving vehicle is known to the transmitting vehicle (obtained via periodic beacons), the inequalities presented in subsection~\ref{subsect:drivinglane} and the relative location information can be used to determine whether human body shadowing is likely in effect. Note that this can change over time for the same receiving vehicle as its relative position to the transmitting vehicle changes. When additional path loss due to human body shadowing is anticipated, the transmitter can then select a lower transmission rate or higher transmission power to compensate and to ensure reliable transmission over the link.
\item \textbf{Multi-hop transmission involving multiple links:} Existing routing schemes~\cite{GPSR,LinkMetric} select the next hop vehicle based on the geometrical distance as well as the link quality to that vehicle. Similar to the case of single link, it can be determined whether human body shadowing is likely in effect. When this is the case, the metric of the packet forwarding scheme for the shadowed link should account for the shadowing to reduce the probability of selecting it.
\item \textbf{Broadcast:} The broadcasting vehicle does not know which vehicle(s) would receive the transmission, and thus cannot use the receiver's relative location to determine whether human body shadowing is in effect.
In this case, the transmitter cannot selectively compensate for the additional path loss due to human body shadowing. However, the application using scooter-to-X communications (i.e., where either the transmitter or the receiver is a scooter) can take into account the reduced link performance on average, including the reduced maximum operation range and lower packet reception probability. This way, it can indirectly compensate for the shadowing effect by, for example, increasing broadcasting frequency.
\end{itemize}
}


\section{Conclusions}
\label{sec:conclusion}

In this paper, we carried out extensive measurements to characterize the scooter-to-X links. The experimental results present challenging channel characteristics \textit{unique to scooters} caused by \textit{human body shadowing}. In cases where the human bodies of the driver and the passenger block the line-of-sight, we observed significant additional signal attenuation ranging from \textit{9 to 18 dB} on average. As an attempt to eliminate human body shadowing, we also consider two dual antenna configurations; however, our study reveals that neither is practical, as they provide limited improvement at significantly higher cost. Furthermore, our simulation results show that, when human body shadowing is not considered, a performance evaluation study could be too optimistic for scooter-to-X links, with reported results of packet delivery rates that are up to 40 percentage points higher than that in a realistic setting. This confirms that incorporating the human body shadowing effect is imperative for scooter-to-X links.

Our study produces several important guidelines for the design of scooter-to-X communication systems. The antenna should be installed at the \textit{highest} possible location on a scooter, in order to avoid the shadowing effect caused by human body as well as the metal body frames of the scooter and other vehicles. This is confirmed by our experimental results. Horizontally, the antenna should be installed \textit{as far from the center of the scooter as possible}, so that, at least for cases with a small transmitter-receiver distance, human body shadowing effect can be mitigated. Based on these observations, we propose a set of inequalities that determine whether human body shadowing would be in effect given the relative locations of the transmitting and receiving vehicles. These results can be utilized in the protocol design, so that shadowed communication links can be avoided when possible, thus increasing the communication reliability of the system. 




\section*{Acknowledgement}
The authors would like thank Tsai-Chih Luo and Shuang-Cheng Yang for their helps in carrying out the additional experiments for the revised manuscript. This work was supported in part by Ministry of Science and Technology of Taiwan, National Taiwan University, and Intel Corporation under grants MOST 103-2911-I-002-001, NTU-ICRP-104R7501, and NTU-ICRP-104R7501-1, and the Croatian Science Foundation under the project IP-2014-09-3877.

\bibliographystyle{elsarticle-num}
\bibliography{Scooter_to_X.bbl}






\newpage
\parpic{\includegraphics[width=1in,height=1.25in,clip,keepaspectratio]{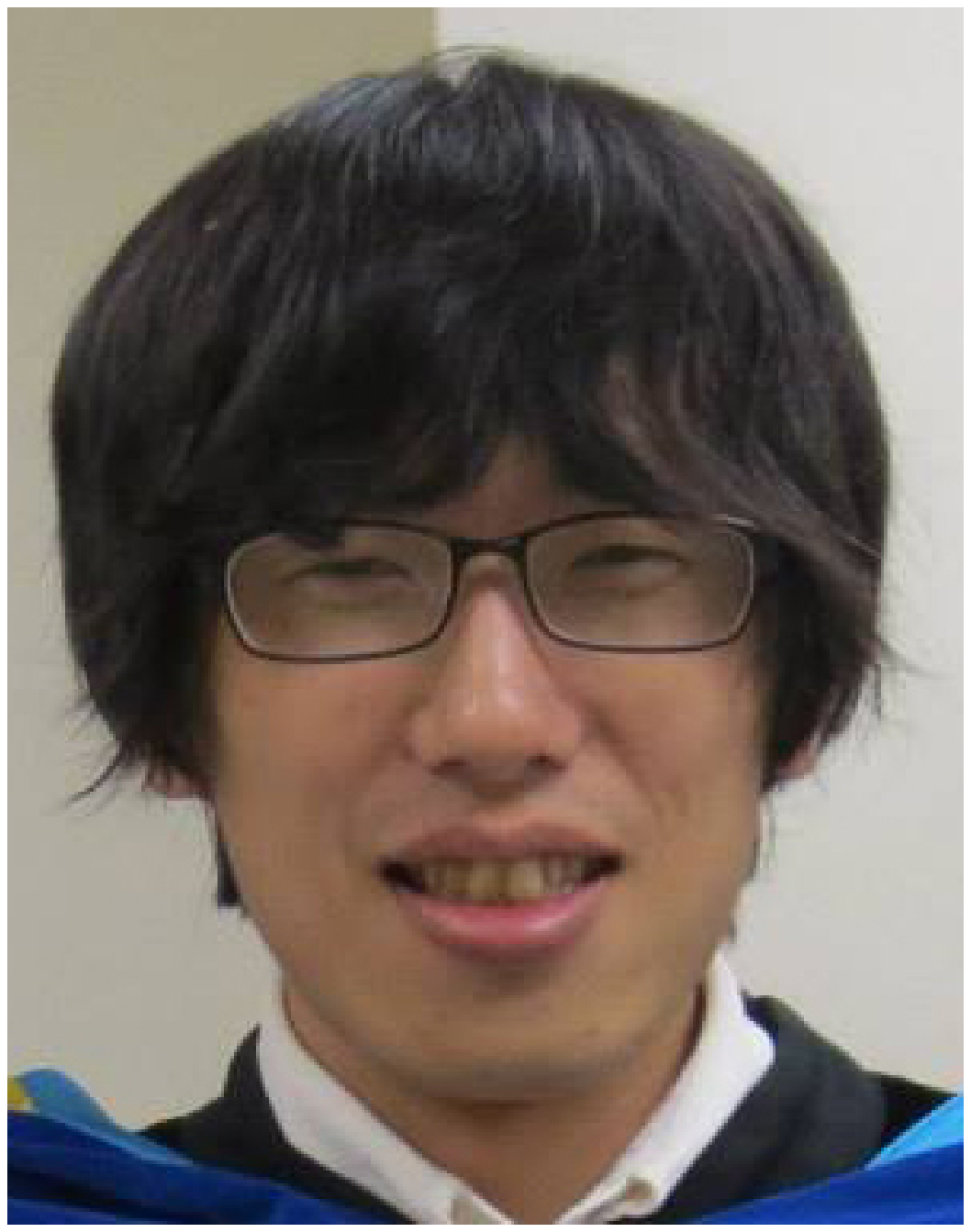}}
\noindent {\bf Hao-Min Lin}
is currently the Deputy Manager of creative software solutions at Moremote, Inc. He received the B.S. degree from Ming Chuan University, M.S. degree from National Ilan University, and his Ph.D. from National Taiwan University, all in computer science and information engineering. His research interests include vehicular ad hoc networks, visible light communications, protocol design, modulation design, and cooperative communications.

\vspace{0.25in}

\parpic{\includegraphics[width=1in,height=1.25in,clip,keepaspectratio]{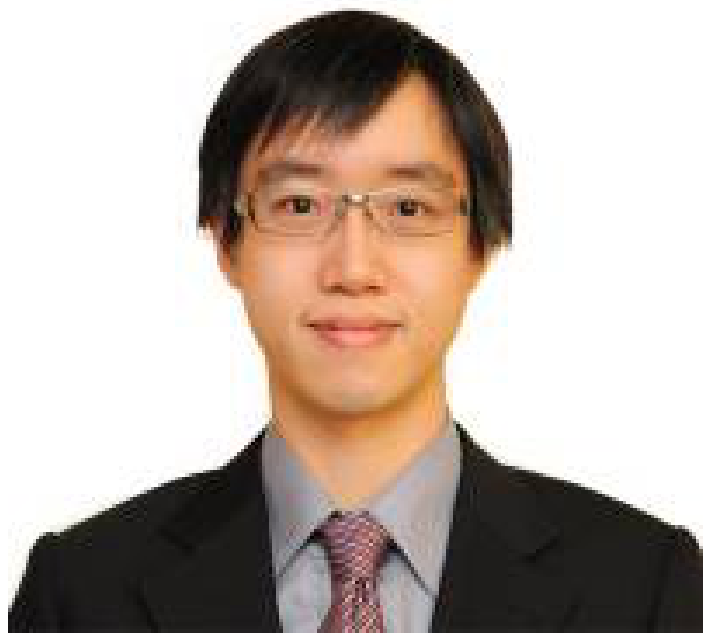}}
\noindent {\bf Hsin-Mu Tsai} is an assistant professor in Department of Computer Science and Information Engineering and Graduate Institute of Networking and Multimedia at National Taiwan University. He received his B.S.E in computer science and information engineering from National Taiwan University in 2002, and his M.S. and Ph.D. in electrical and computer engineering from Carnegie Mellon University in 2006 and 2010, respectively. Dr. Tsai's recognitions include 2013 Intel Early Career Faculty Award (the first to receive this honor outside of North America and Europe), 2014 Intel Labs Distinguished Collaborative Research Award, and National Taiwan University's Distinguished Teaching Award. He served as TPC co-chair for ACM VANET 2013 and the workshop co-chair for the first ACM Visible Light Communication System (VLCS) Workshop (2014). His research interests include vehicular networking and communications, wireless channel and link measurements, vehicle safety systems, and visible light communications.

\vspace{0.25in}

\parpic{\includegraphics[width=1in,height=1.25in,clip,keepaspectratio]{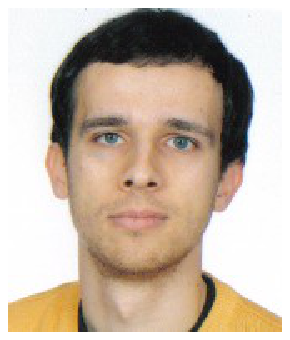}}
\noindent {\bf Mate Boban} is a senior researcher at Huawei European Research Center, Munich. He earned his Ph.D. degree in electrical and computer engineering from Carnegie Mellon University and his diploma in Informatics from University of Zagreb. Before joining Huawei, he worked for NEC Laboratories Europe, Carnegie Mellon University, and Apple Inc. He is an alumnus of the Fulbright Scholar Program. His current research is in the areas of intelligent transportation systems, wireless communications, and networking. He received the Best Paper Award at the IEEE VTC Spring 2014 and at IEEE VNC 2014.

\end{document}